\documentclass[amsmath,amssymb,aps,pra,reprint,superscriptaddress,longbibliography]{revtex4-2}

\usepackage[svgnames]{xcolor}
\usepackage{chemformula} 
\usepackage{subfig}
\usepackage{tabularx}
\usepackage{url}
\usepackage{cancel}
\usepackage{quantikz}
\usepackage[colorlinks,citecolor=DarkGreen,linkcolor=FireBrick,linktocpage,unicode,hypertexnames=false]{hyperref}

\begin{document}

\author{Viktor Khinevich}%
\email{victorkh711@gmail.com}
\affiliation{%
  Graduate School of Engineering Science, The University of Osaka, 1-3 Machikaneyama, Toyonaka, Osaka 560-8531, Japan
}%
\affiliation{%
  Center for Quantum Information and Quantum Biology,
  The University of Osaka, 1-2 Machikaneyama, Toyonaka 560-8531, Japan
}%

\author{Wataru Mizukami}%
\email{mizukami.wataru.qiqb@osaka-u.ac.jp}
\affiliation{%
  Graduate School of Engineering Science, The University of Osaka, 1-3 Machikaneyama, Toyonaka, Osaka 560-8531, Japan
}%
\affiliation{%
  Center for Quantum Information and Quantum Biology,
  The University of Osaka, 1-2 Machikaneyama, Toyonaka 560-8531, Japan
}%

\title{Enhancing quantum computations with the synergy of auxiliary field quantum Monte Carlo and computational basis tomography}

\begin{abstract}
We introduce QC-CBT-AFQMC, a hybrid algorithm that incorporates computational basis tomography (CBT) into the quantum-classical auxiliary-field quantum Monte Carlo (QC-AFQMC) method proposed by Huggins et al. [{\textit{Nature}} \textbf{603}, 416-420 (2022)], replacing the use of classical shadows. While the original QC-AFQMC showed high accuracy for quantum chemistry calculations, it required exponentially costly post-processing.
Subsequent work using Matchgate shadows [{\textit{Commun. Math. Phys.}} \textbf{404}, 629 (2023)] improved scalability, but still suffers from prohibitive computational requirements that limit practical applications.
Our QC-CBT-AFQMC approach uses shallow Clifford circuits with a quadratic reduction of two-qubit gates over the original algorithm, significantly reducing computational requirements and enabling accurate calculations under limited measurement budgets. We demonstrate its effectiveness on the hydroxyl radical, ethylene, and nitrogen molecule, producing potential energy curves that closely match established benchmarks. We also examine the influence of CBT measurement counts on accuracy, showing that subtracting the active space AFQMC energy mitigates measurement-induced errors. Furthermore, we apply QC-CBT-AFQMC to estimate reaction barriers in [3+2]-cycloaddition reactions, achieving agreement with high-level references and successfully incorporating complete basis set extrapolation techniques.
These results highlight QC-CBT-AFQMC as a practical quantum-classical hybrid method that bridges the capabilities of quantum devices and accurate chemical simulations.
\end{abstract}

\maketitle

\section{Introduction}
Over the past decade, quantum computing has significantly advanced, particularly in quantum chemistry \cite{Bauer2020,Motta_2021}. 
Many quantum algorithms for quantum chemistry have been proposed, among which the variational quantum eigensolver (VQE)~\cite{Peruzzo2014}, and quantum phase estimation (QPE)~\cite{Lloyd1996,Aspuru2005}, are widely recognized as representative methods. Nonetheless, existing algorithms, including both VQE and QPE, are resource-intensive, making their application to full electronic structure Hamiltonians impractical. 

To address this, the active space approach from classical computational chemistry is essential, which involves selecting critical molecular orbitals and electrons, thus focusing on static electron correlation. 
Quantum computers efficiently handle rapidly increasing configurations within the active space. However, capturing the dynamical correlation outside this space remains challenging. There are various attempts to incorporate dynamical correlation after quantum computing. One of such methods is the multireference perturbation theory~\cite{Krompiec2022, Tammaro2023,Mitarai2023,Cortes2023}. 
However, this requires heavy measurements of $3$,$4$-reduced density matrices, or other quantities~\cite{Nishio2023}.
Other approaches involve explicitly correlated methods, such as F$12$ and transcorrelation methods \cite{Schleich2022, McArdle2020,Kumar2022,Sokolov2023}.
For F$12$ methods, the inclusion of the Coulomb cusp condition alone is insufficient to adequately capture dynamical correlation, while the transcorrelation methods are limited by the emergence of three-body terms and loss of hermicity, creating practical implementation bottlenecks.

\begin{figure*}[!t]
\includegraphics[width=0.70\textwidth]{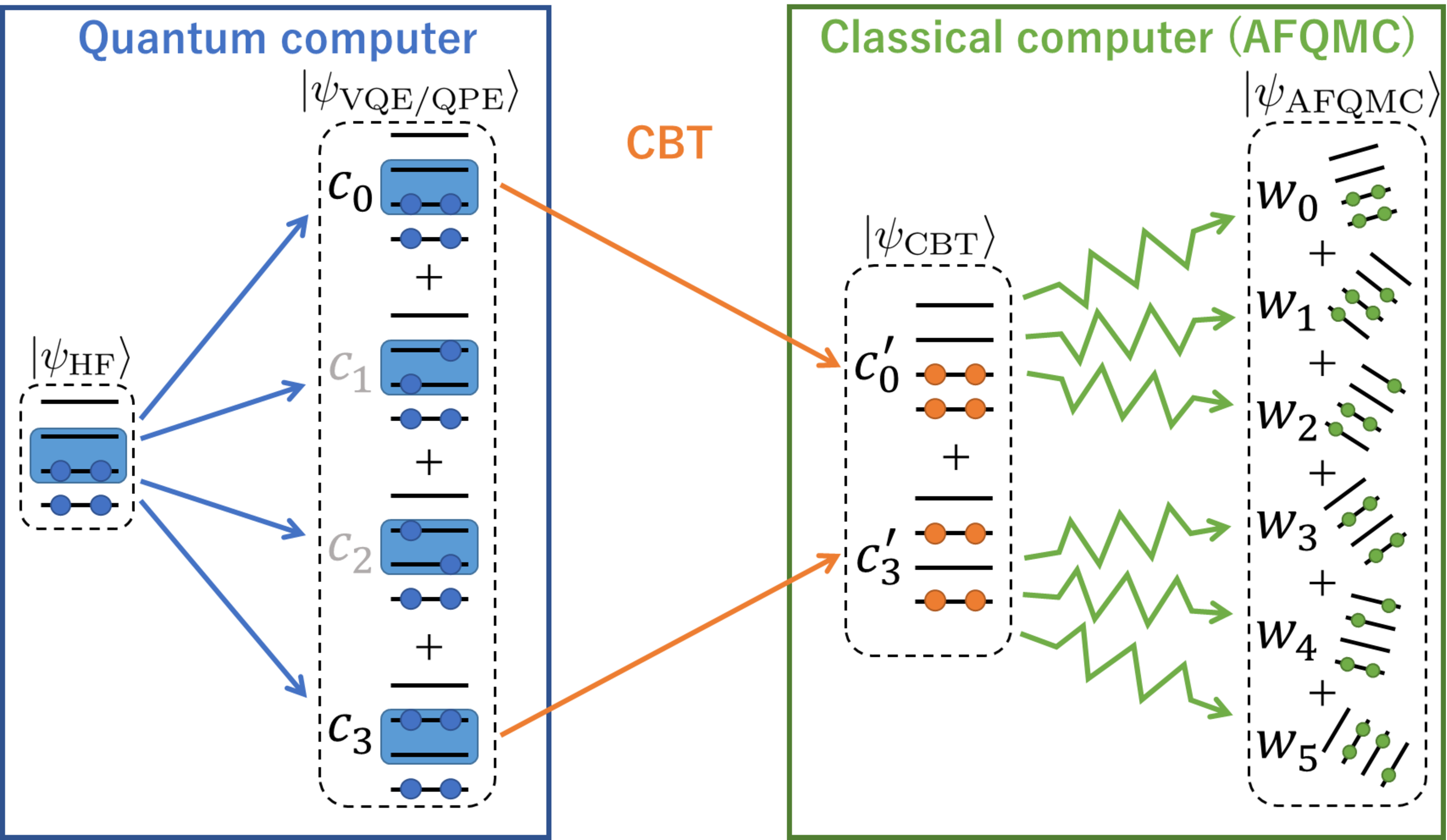} 
\caption{Schematic of the QC-CBT-AFQMC methodology. Blue frames highlight the selected active space for the quantum computer part. The CBT algorithm captures the most important configurations from the quantum computer state and translates them to the classical computer. AFQMC generates walkers from the CBT state, where the orbitals also rotate. This is depicted as rotated orbital diagrams.} \label{Figure_scheme_CBT_AFQMC} 
\end{figure*}

Huggins et al. proposed a quantum-classical hybrid algorithm, named QC-AFQMC~\cite{Huggins2022}, where a quantum-prepared trial wave function is used to reduce bias in phaseless auxiliary-field quantum Monte Carlo (ph-AFQMC)~\cite{Zhang2003}. Although QC-AFQMC originally introduced to improve a AFQMC calculation with quantum computing, it can also be viewed as a post-processing scheme that describes residual dynamical correlation beyond the active space~\cite{Zhao2025arXiv,Goings2025arXiv}.
In the framework of QC-AFQMC, a quantum computer is used to prepare a trial wave function, employed to handle the infamous sign problem, for the AFQMC method. A significant advantage of AFQMC is its scaling $O(N^3)-O(N^4)$ with respect to the system size $N$~\cite{Motta2018,Lee2022}, which is close to Hartree-Fock. The favorable scaling ensures that AFQMC remains computationally tractable for larger systems. 
These advantages have led AFQMC methods to become increasingly popular for capturing dynamic correlation effects~\cite{Chen2023, yoshida2025auxiliaryfieldquantummontecarlo, kiser2024contextualsubspaceauxiliaryfieldquantum}.
In the original QC-AFQMC paper, the translation of the quantum wave function into its classical counterpart was achieved using Clifford shadow tomography \cite{Huang2020}. It was found that QC-AFQMC is inherently noise resilient, since the Monte Carlo step effectively corrected the errors introduced by the noisy quantum step. However, this method requires one to use the postprocessing with exponential scaling on classical computers. 

In order to overcome this bottleneck, the Matchgate shadow tomography technique was used~\cite{Wan2023}. Although this approach allows to reduce the scaling to polynomial, the degree of this polynomial is much higher than that of AFQMC itself \cite{huang2024evaluatingquantumclassicalquantummonte}. The most expensive operation is to compute the local energy for each walker at each time step, which scales as $O(N_q^{8.5} \log^2(N_q))$ with the number of qubits $N_q$. Another problem of this approach is the large constant prefactor, due to the high sampling requirements. When specific physical quantities or information are targeted, this method may not be the optimal choice in terms of measurement count due to shot budget constraints~\cite{takemori2023balancing}. Also, it was previously noted that all implementations of QC-AFQMC may require an exponential number of shots due to the exponential decay of the overlap of the trial state with the walkers \cite{mazzola2022exponentialchallengesunbiasingquantum, lee2022responseexponentialchallengesunbiasing}. 

To overcome these limitations of QC-AFQMC for practical calculations, we introduce the QC-CBT-AFQMC method, which applies computational basis tomography (CBT) \cite{Kohda2022} to QC-AFQMC as a technique to extract wave function information from quantum computers in an efficient way to use on classical computers. The practicality of our approach comes from a lower requirement for the number of shots. A smaller constant prefactor and better scaling $O(N_q^4)$ than for Matchgates allowed us to do calculations for much larger systems than previously. Also, CBT utilizes shallow circuits, built from only Clifford gates; our method can be implemented even on near-term quantum devices. CBT circuits require $O(N_q)$ number of CNOT gates. The depth of the CBT circuit can also be reduced and even made constant due to the properties of the fan-out gates \cite{baumer2024measurementbasedlongrangeentanglinggates}.
Using CBT, we have already succeeded in combining coupled cluster and quantum computing through a method called tailored coupled cluster~\cite{Erhart2024}.

The structure of this paper is as follows.
Section~II summarizes the CBT and outlines the integration of CBT into the QC-AFQMC workflow. 
Section~III details our numerical experiments, examining potential energy curves and the effect of the shot budget on accuracy. 
Finally, Section~IV concludes with an outlook on further improvements and possible expansions of this methodology, including the incorporation of alternative approaches.
\section{Methods}

\subsection{Computational Basis Tomography}

\begin{figure}[!ht]
  \centering
  \begin{quantikz}[column sep=0.5cm]
        \lstick{$q_0$} & \targ{}\gategroup[4,steps=3,style={dashed,rounded
            corners,fill=blue!20, inner
            xsep=2pt},background,label style={label
            position=above,anchor=north,yshift=+0.2cm}]{{fan-out gate}}
        & \qw      & \qw       & \qw       & \qw      & \qw \\
        \lstick{$q_1$} & \qw          & \targ{}      & \qw       & \qw       & \qw      & \qw \\
        \lstick{$q_2$} & \ctrl{-2}    & \ctrl{-1}      & \ctrl{1}  & \gate{S}\gategroup[1,steps=1,style={dashed,rounded
        corners,fill=red!20, inner
        xsep=2pt},background,label style={label
        position=above,anchor=north,yshift=+0.2cm}]{{}} 
        & \gate{H} & \qw \\
        \lstick{$q_3$} & \qw          & \qw      & \targ{}   & \gate{X}  & \qw      & \qw 
  \end{quantikz}
  \caption{Quantum circuit implementing the operators $\langle0000|U_{1,14} = (\langle0001| + \langle1110|)/\sqrt{2}$, and $\langle0000|V_{1,14} = (\langle0001| + i\langle1110|)/\sqrt{2}$. The blue dashed box indicates a fan-out gate from the flag qubit $q_2$ to the other qubits, corresponting to different bits in binary representations of $1$ and $14$. The additional $S$ gate transforms $U_{1,14}$ into $V_{1,14}$.}
  \label{fig:CBT_circuit}
\end{figure}

In quantum computing, a wave function is typically represented as a linear combination of basis functions. An arbitrary wave function can be decomposed into a series of bit strings of length $N$ as follows:
\begin{equation} \label{cbt_1}
    |\psi\rangle = \sum^{2^N-1}_{n=0}\langle n|\psi\rangle|n\rangle  
\end{equation}

However, extracting the coefficients $\langle n|\psi\rangle$ from a quantum device poses a challenge, as they are encoded as quantum information. The retrieval of such data for use in hybrid quantum-classical algorithms requires measurement. To do this, we utilized an efficient measurement strategy known as CBT.

The CBT technique was designed to extract classical information about a wave function that is encoded on a quantum device. This extraction is achieved through projection measurements executed within the computational basis. The foundation of CBT lies in the computational basis sampling method, which was originally used for estimating expectation values, as described previously \cite{Kohda2022}. CBT can extract CI coefficients using shallow Clifford circuits, making it appealing for systems where shot budgets are limited. Although CBT may particularly benefit states where only a limited number of configurations dominate, our primary motivation is its practical feasibility: it requires fewer two-qubit gates and less circuit depth and allows efficient coefficient extraction even within the constraints of near-term quantum devices.

The initial phase of CBT involves determining the absolute magnitudes of the coefficients $\langle n|\psi\rangle$ shown in equation \ref{cbt_1}. To facilitate this, it is necessary to generate $N_f$ identical instances of the wave function $|\psi \rangle$ and perform measurements within the computational basis, which encompasses all possible bit strings ${|n\rangle}^{2^N-1}_{n=0}$ of length $N$. Consequently, the absolute values of the coefficients are estimated using the formula:
\begin{equation} \label{cbt_2}
    |\langle n|\psi\rangle| \approx \sqrt{\frac{N_n}{N_f}},
\end{equation}
where $N_n$ denotes the frequency of observing the specific bit string $n$.

After these are measured, we select several of the most significant bit strings in the decomposition \ref{cbt_1}. We assume that, if we choose an appropriate single-particle basis, there is a relatively small set of configurations in chemistry-related systems that carry non-negligible weight, specifically those that do not exhibit exponential or high-polynomial decay with respect to system size. Given this assumption, we choose to keep only the $R$ most probable outcomes; our method does not work when this assumption breaks down, say, for quantum states with uniformly exponentially decaying CI coefficients. This step reduces the number of necessary measurements while preserving the accuracy of the wave function. 

At this point, the absolute magnitudes of the $R$ chosen coefficients $\langle n|\psi\rangle$ have been determined. Nevertheless, this information alone is insufficient for reconstructing the wave function due to the loss of phase information. To recover the phase details, we employed the following equation:
\begin{equation} \label{cbt_3}
    e^{i(\phi_n-\phi_m)}=\frac{\langle n|\psi \rangle \langle \psi|m\rangle}{|{\langle n|\psi \rangle}||{\langle \psi|m\rangle}|}.
\end{equation}
It is important to note that phase determination is achievable only up to an arbitrary global phase. Thus, the phase $\phi_m$ was arbitrarily set to zero, which allows the calculation of all other phases relative to $\phi_m$.

To utilize equation \ref{cbt_3}, it is necessary to compute the value of $\langle n|\psi \rangle \langle \psi|m\rangle$, referred to as the \textit{interference factor}, with the denominator already determined from the previous step. The auxiliary relationship was determined as follows:
\begin{equation} \label{cbt_4}
    \langle n|\psi \rangle \langle \psi|m\rangle = \mathcal{A}_{n,m} + i \mathcal{B}_{n,m} - \frac{1+i}2 (|{\langle n|\psi \rangle}|^2 + |{\langle m|\psi\rangle}|^2),
\end{equation}
where the terms $\mathcal{A}_{n,m}$ and $\mathcal{B}_{n,m}$ are defined by the equations
\begin{equation} \label{cbt_5}
    \mathcal{A}_{n,m} \equiv |{\frac{\langle n| + \langle m|}{\sqrt{2}}|\psi \rangle}|^2 
    ,\,
    \mathcal{B}_{n,m} \equiv |{\frac{\langle n| + i\langle m|}{\sqrt{2}}|\psi\rangle}|^2.
\end{equation}
These quantities represent the only unknowns in equation \ref{cbt_4}. Ancillary unitary operators $U_{n,m}$ and $V_{n,m}$ were constructed as described in \cite{Kohda2022}: $U^\dagger_{m,n}|0\rangle= (|m\rangle + |n\rangle)/\sqrt{2}$ and $V^\dagger_{m,n}|0\rangle= (|m\rangle - i |n\rangle)/\sqrt{2}$. These operators can be written in Clifford circuits, each of which requires only $\operatorname{Hamming}(m,n)$ of CNOT gates. The construction proceeds by identifying the qubit position at which the binary representations of $m$ and $n$ differ to use it as a flag and applying a Hadamard gate there. Then, for each remaining bit that differs, a CNOT gate is applied with the control on the flag qubit and the target on the differing bit. To construct $V_{m,n}$ instead of $U_{m,n}$, an additional $S$ gate is applied to introduce the required relative phase. An example of this construction for $m=1$ and $n=14$ is shown in Figure~\ref{fig:CBT_circuit}. This approach guarantees that each such unitary requires at most $O(N_q)$ CNOT gates and CNOT layers, whereas arbitrary Matchgate circuits typically require $O(N_q^2)$ gates. Also, these CNOT gates used in $U_{n,m}$ and $V_{n,m}$ comprise fan-out gates, allowing to implement it with constant CNOT layers using mid-circuit measurements \cite{baumer2024measurementbasedlongrangeentanglinggates}. The operators $U_{n,m}$ and $V_{n,m}$ were incorporated to redefine $\mathcal{A}_{n,m}$ and $\mathcal{B}_{n,m}$ as
\begin{equation} \label{cbt_6}
    \mathcal{A}_{n,m} = |{\langle 0|U_{n,m}|\psi \rangle}|^2 
    ,\,
    \mathcal{B}_{n,m} = |{\langle 0|V_{n,m}|\psi\rangle}|^2.
\end{equation}
From equations \ref{cbt_6}, it is evident that the values of $\mathcal{A}_{n,m}$ and $\mathcal{B}_{n,m}$ can be derived from measurements on the zero state $|0\rangle$. Consequently, to evaluate $\mathcal{A}_{n,m}$ and $\mathcal{B}_{n,m}$ for each distinct bit string pair $n \neq m$, $N_a$ and $N_b$ measurements must be performed, respectively, yielding: 
\begin{equation}
    \mathcal{A}_{n,m} \approx \sqrt{\frac{N^{(a)}_0}{N_a}} 
    ,\,
    \mathcal{B}_{n,m} \approx \sqrt{\frac{N^{(b)}_0}{N_b}},
\end{equation}
where $N^{(a)}_0$ and $N^{(b)}_0$ indicate the counts of zero outcomes for the respective measurement sets. With these estimates, \ref{cbt_4} is applied to determine the phases for all coefficients $\langle n|\psi\rangle$. Thus, the complete CBT protocol necessitates a total of $N_f + (R-1)(N_a + N_b)$ measurements of the state $|\psi \rangle$, where $R$ is the number of selected bit strings for the correct representation of the wave function.
The CBT wave function has the following form:
\begin{equation}
\label{CBT_WF}
     |\psi_\text{CBT} \rangle = \sum^{R}_{n=0} c'_n |n\rangle,
\end{equation}
where $c'_n = |\langle n | \psi \rangle| e^{i (\phi_n - \phi_0)}$ are normalized coefficients, obtained from equations \ref{cbt_2} and \ref{cbt_3}. Both the absolute values and phases are estimated from sampling results, and therefore carry statistical errors.

\subsection{Auxiliary Field Quantum Monte Carlo Method}

The AFQMC method is a state-of-the-art approach for obtaining high-precision solutions to the time-independent Schrödinger equation \cite{Sugiyama1986}. It offers computational advantages over both deterministic and stochastic high-accuracy methods. AFQMC retains the low, Hartree-–Fock–like scaling of $O(N^3\!-\!N^4)$, yet delivers precision comparable to high-precision methods, such as full CI, multireference perturbation theory and coupled cluster. It also has advantages over diffusion Monte Carlo (DMC) and full configuration interaction QMC (FCIQMC), since its use of a Slater determinant trial wave function that controls the sign problem and can be efficiently processed on GPU or multicore CPU architectures \cite{Malone2022}.

In AFQMC, the ground‐state wave function is obtained by evolving an initial state $|\psi(0)\rangle$ in imaginary time:
\begin{equation} 
\label{afqmc_1} 
    -\frac{d}{d\tau} |\psi(\tau)\rangle = (\hat{H} - E_0)|\psi(\tau)\rangle, 
\end{equation} 
where $\tau=it$ is the imaginary time and $E_0$ is the energy of the ground state. The solution is as follows: 
\begin{equation} 
\label{afqmc_2} 
    |\psi(\tau)\rangle = e^{-\tau(\hat{H} - E_0)}|\psi(0)\rangle.
\end{equation} 

As $\tau \to \infty$, $\ket{\psi(\tau)}$ is projected onto the ground state, provided $E_0$ and $|\psi(0)\rangle$ are accurately estimated. In practice, $E_0$ is iteratively refined alongside updates to the wave function.

To compute the propagation in Equation \ref{afqmc_2}, time discretization is used: 
\begin{equation} 
\label{afqmc_3} 
    e^{-\tau (\hat{H} - E_0)} = \left( e^{-\Delta \tau (\hat{H} - E_0)} \right)^n, 
\end{equation} 
with $\Delta \tau = \tau / n$.

Direct computation is as challenging as solving the full configuration interaction problem. AFQMC circumvents this by using a Hubbard–Stratonovich transformation to decouple the two‐body interactions into one‐body operators coupled to stochastic auxiliary fields. It converts the interacting system into non-interacting particles under a fluctuating external potential.

To apply this transformation, one first expresses the Hamiltonian in quadratic form via the Cholesky decomposition: 
\begin{equation} 
\label{afqmc_4} 
    \hat{H} = \hat{\nu}_0 - \frac{1}{2} \sum_{i=1}^{N_{\text{max}}} \hat{\nu}_i^2, 
\end{equation} 
where 
\begin{align*} 
    \hat{\nu}_0 = \sum_{pq} h_{pq} \sum_{\sigma} \hat{a}_{p\sigma}^{\dagger} \hat{a}_{q\sigma}, 
    \hat{\nu}_i = \mathrm{i} \sum_{pq} L_{pq}^i \sum_{\sigma} \hat{a}_{p\sigma}^{\dagger} \hat{a}_{q\sigma}.
\end{align*} 
Here, $\hat{a}_{p\sigma}^\dagger$ and $\hat{a}_{q\sigma}$ are the fermionic creation and annihilation operators for an electron in spatial orbital $p$ or $q$ with spin projection $\sigma$. The indices $p$ and $q$ label a set of orthonormal single-particle spatial orbitals. The coefficients $h_{pq}$ are the matrix elements of the one-electron part of the Hamiltonian. The quantities $L_{pq}^i$ are obtained from the Cholesky decomposition of the tensor of the two-electron repulsion integrals $V_{pqrs}$, approximated as $V_{pqrs} \approx \sum_i L_{pr}^i L_{qs}^i$.

Since $\hat{\nu}_0$ and $\hat{\nu}_i$ generally do not commute, the first-order Trotter decomposition is used: 
\begin{equation} 
\label{afqmc_5} 
    e^{-\Delta \tau (\hat{H} - E_0)} \approx e^{-\frac{\Delta \tau}{2} \hat{\nu}_0} \prod_{i} e^{\frac{\Delta \tau}{2} \hat{\nu}_i^2} e^{-\frac{\Delta \tau}{2} \hat{\nu}_0}. 
\end{equation}

Now, the propagator is ready to apply the Hubbard-Stratonovich transformation, yielding the following: 
\begin{equation} 
\label{afqmc_6} 
    e^{-\Delta \tau(\hat{H} - E_0)} \approx \int d\vec{x} \, p(\vec{x}) \hat{B}(\vec{x}), 
\end{equation} 
where $p(\vec x)$ is a normal distribution over the auxiliary field $\vec x$ and $\hat{B}({\vec x}) = \exp\left(-\Delta \tau(\hat{\nu}_0 - E_0) + \sqrt{\Delta \tau}\,\vec{x}\cdot\hat{\vec{\nu}}\right)$ is the evolution of a single particle. Here, $\vec{x}$ is an auxiliary field that simulates particle interactions by coupling to this field.

Equation \ref{afqmc_6} allows propagation of the initial wave function, with the integral over $\vec{x}$ evaluated via Monte Carlo techniques. However, direct use faces the sign problem, which can be mitigated by the phaseless approximation (ph-AFQMC) \cite{Zhang2003}.

In ph-AFQMC, an importance function $I(\vec{x}; \psi(\tau))$ is incorporated into the propagator: 
\begin{equation} 
\label{afqmc_7} 
    e^{-\Delta \tau(\hat{H} - E_0)} \approx \int d\vec{x} \, p(\vec{x}) I(\vec{x}; \psi(\tau)) \hat{B}(\vec{x} - \langle \vec{x} \rangle).
\end{equation}

The sign problem arises from the fact that the coefficients $\hat{\nu}_i$ may carry arbitrary complex phases, including negative real values. In the multireference case, the propagator alters the phases of Slater determinants and makes orbitals complex, potentially leading to cancellations in the Monte Carlo averaging and thus causing divergence of the wave function and energy. The phaseless approximation uses an importance function to prevent abrupt phase changes, ensuring that the coefficients remain sufficiently large. A commonly used importance function is: 
\begin{equation} 
\label{afqmc_8} 
    I(\vec{x}; \psi(\tau)) = e^{-\Delta \tau (\mathrm{Re} (E_L) - E_0)} \cdot \max\left( 0, \cos(\Delta \theta) \right), 
\end{equation} where $\Delta \theta = \mathrm{Arg} \left(\frac{\langle \psi_T|\psi_{k+1,w}\rangle}{\langle \psi_T|\psi_{k,w}\rangle}\right)$ represents the change in the argument of the overlap between the trial wave function $|\psi_T\rangle$ and the propagated wave function. The local energy is $E_L(\tau) = \langle \psi_T | \hat{H} | \psi(\tau) \rangle / \langle \psi_T | \psi(\tau) \rangle$. The force bias $\langle \vec x \rangle = - \sqrt{\Delta \tau} \langle \psi_T | \hat{\vec \nu} | \psi(\tau) \rangle / \langle \psi_T | \psi(\tau) \rangle$ minimizes the fluctuations in $I(\vec{x}; \psi(\tau))$. The trial wave function $|\psi_T\rangle$ is typically the same as the initial guess but can differ.

The evolution of Slater determinants and their weights are described by: 
\begin{equation} 
\label{afqmc_9} 
    \begin{split} 
    |\psi_{k+1,w}\rangle &= \hat{B}(\vec{x}_{k,w} - \langle \vec{x}_{k,w} \rangle) |\psi_{k,w}\rangle, \\
    W_{k+1,w} &= I(\vec{x}_{k,w}; \psi_{k,w}) W_{k,w}, 
    \end{split}
\end{equation} 
where $k$ is the time step and $w$ indexes the Slater determinants.

The energy is calculated using: 
\begin{equation} 
\label{afqmc_10} 
    E(\tau) = \frac{\sum_w W_w(\tau) E_L(\tau)}{\sum_w W_w(\tau)},
\end{equation} 
with $W_w(\tau) = W_{w,k}$ at $\tau = k \Delta \tau$.

\subsection{QC-CBT-AFQMC}

The essential features of QC-CBT-AFQMC are depicted in Figure \ref{Figure_scheme_CBT_AFQMC}. The initial phase of the QC-CBT-AFQMC method involves executing a quantum algorithm, such as VQE or QPE, within a predefined active space. This step is crucial for capturing the electron correlation within the selected active space, which predominantly consists of static correlation in systems that require multireference approaches. This phase yields estimates of the energy $E^{\text{act}}_{\text{VQE/QPE}}$ and the initial wave function $|\psi_{\text{VQE/QPE}}\rangle$. However, this wave function, encapsulated as quantum information, is not directly applicable to subsequent AFQMC calculations.

CBT is used to convert the acquired wave function for use in the classical part of the algorithm. By preparing the state $|\psi_{\text{VQE/QPE}}\rangle$ and conducting measurements as outlined in the CBT framework, the CI coefficients are reconstructed, and the $|\psi_{\text{CBT}}\rangle$ wave function is formulated. CBT works under the assumption that a relatively small set of configurations have reasonably large coefficients. If, despite orbital optimization, all coefficients were uniformly exponentially decaying, CBT would break down; however, in such a pathological regime, any tomography-based QC-AFQMC scheme would likewise fail, as no meaningful multi-determinant trial state could be constructed. The CBT wave function serves as the initial guess for the AFQMC computation. It is important to note that the $|\psi_{\text{CBT}}\rangle$ wave function now incorporates errors originating from both the VQE/QPE and CBT processes because in practice, the number of measurements and described configurations of $|\psi_{\text{VQE/QPE}}\rangle$ is limited.

An important advantage of QC-CBT-AFQMC over the Matchgate shadows approach \cite{Wan2023} lies in the calculation of the local energy $E_L(\tau)$ and the force bias $\langle \vec{x} \rangle$ (Equations \ref{afqmc_8} and \ref{afqmc_10}). In our method, these quantities are evaluated using the classically reconstructed multi-determinant trial wave function $|\psi_{\text{CBT}}\rangle$, in exactly the same way as conventional ph-AFQMC with multi-Slater determinants as a trial wave function. The necessary overlaps and Hamiltonian matrix elements between $|\psi_{\text{CBT}}\rangle$ and the Slater determinant walkers can be efficiently computed, scaling polynomially with the number of qubits (active space size) $N_q$ and linearly with the number of selected determinants $R$. As a result, the classical computational cost within the AFQMC loop remains dominated by standard steps that scale as $O(N_q^4)$, consistent with conventional AFQMC methods. In contrast, for Matchgate shadows, the local energy must be repeatedly estimated for each walker at each time step, and this procedure is the most computationally demanding part of the approach, scaling as $O(N_q^{8.5} \log^2(N_q))$ with active space size \cite{huang2024evaluatingquantumclassicalquantummonte}. Moreover, CBT allows for accurate computation and storage of amplitudes and their ratios, even for truncated trial wave functions, which is particularly important for mitigating the problem of exponential decay in the overlap between the trial wave function and AFQMC walkers. This issue otherwise leads to an exponential number of shots required to reconstruct the trial wave function \cite{mazzola2022exponentialchallengesunbiasingquantum, lee2022responseexponentialchallengesunbiasing}.

Once the $|\psi_{\text{CBT}}\rangle$ wave function is reconstructed, it is used as a trial wave function for the AFQMC calculation. No further CBT is performed, and all subsequent calculations are carried out on a classical computer. AFQMC is used to address the dynamical electron correlation missing from the active space. We assume that the errors from the CBT measurements are larger than those from the VQE or QPE steps. To isolate the correlation energy external to the active space, the energy computed using the full electronic Hamiltonian via AFQMC, $E^{\text{Full}}_{\text{AFQMC}}$, is adjusted by subtracting the energy determined using the active space Hamiltonian, $E^{\text{act}}_{\text{AFQMC}}$. Therefore, for each trial wave function $|\psi_{\text{CBT}}\rangle$, two distinct AFQMC evaluations are conducted.

Upon gathering all necessary data, the overall energy within the QC-CBT-AFQMC framework, $E_{\text{QC-CBT-AFQMC}}$, is calculated as follows:
\begin{equation} \label{cbt-afqmc_1}
    E_{\text{QC-CBT-AFQMC}} = E^{\text{act}}_{\text{VQE/QPE}} + (E^{\text{Full}}_{\text{AFQMC}} - E^{\text{act}}_{\text{AFQMC}}). 
\end{equation}
This equation assumes that errors from CBT measurements will cancel out, while the correlation energy external to the active space is cumulatively added to $E^{\text{act}}_{\text{VQE/QPE}}$. Subsequent analysis results demonstrate that this assumption is true, especially when CBT-induced errors are greater than those from the VQE/QPE stages.
Virtually the same correction was used in our tailored coupled cluster study with CBT~\cite{Erhart2024}.
\section{Results and Discussion}

This section demonstrates the efficacy of the proposed method for constructing potential energy curves (PECs) of selected molecular systems, including the hydroxyl radical, ethylene, and nitrogen molecules. We compare the PECs generated via QC-CBT-AFQMC with those obtained from established computational techniques such as multireference configuration interaction, multireference perturbation theory, and coupled cluster methods. Furthermore, the impact of the number of CBT measurements on the accuracy of the results is evaluated, both with and without the application of equation \ref{cbt-afqmc_1}. Additionally, the reaction barriers for two [$3$+$2$] cycloaddition reactions are evaluated, and the application of CBS extrapolation techniques to enhance QC-CBT-AFQMC calculations is discussed. Although a detailed resource analysis of Matchgate-shadow-based QC-AFQMC has recently been published \cite{huang2024evaluatingquantumclassicalquantummonte}, the corresponding implementation has not yet been released in an open source repository. Since developing one would require considerable additional effort, we did not include the computational results of this method as a direct reference in the present study.

The initiation of the CBT-AFQMC computational workflow involved executing the VQE using the unitary coupled cluster (UCCSD) ansatz, facilitated by the Chemqulacs package \cite{chemqulacs}. We assumed that the VQE stage errors are much smaller than those introduced by the CBT measurements. As a result, we performed all VQE calculations in a noiseless setting. Subsequent steps leveraging CBT were developed using the Qulacs framework \cite{Suzuki2021}. For the AFQMC step, the Ipie package was used to ensure efficient quantum Monte Carlo simulations \cite{Malone2022}. Benchmark PECs were generated using various computational approaches, including self-consistent field (SCF), complete active space configuration interaction (CASCI), strongly contracted n-electron valence perturbation theory (SC-NEVPT$2$), and coupled cluster theory (CCSD(T)), all implemented within the PySCF package \cite{Sun2018}. The selection of active spaces and basis sets was tailored specifically to each molecule, as described later in detail. Given the unsatisfactory performance of CCSD(T) PEC for \ch{N2}, multireference configuration interaction (MRCISD+Q) was adopted as a more accurate reference method. Due to the well-documented size-consistency issues of MRCISD, the Davidson correction was applied to mitigate this shortcoming \cite{Langhoff1974}. The MRCISD+Q method was conducted using the \textit{ab initio} Orca package \cite{Neese2012}. 

\begin{figure}[ht]
    \includegraphics[width=0.45\textwidth]    {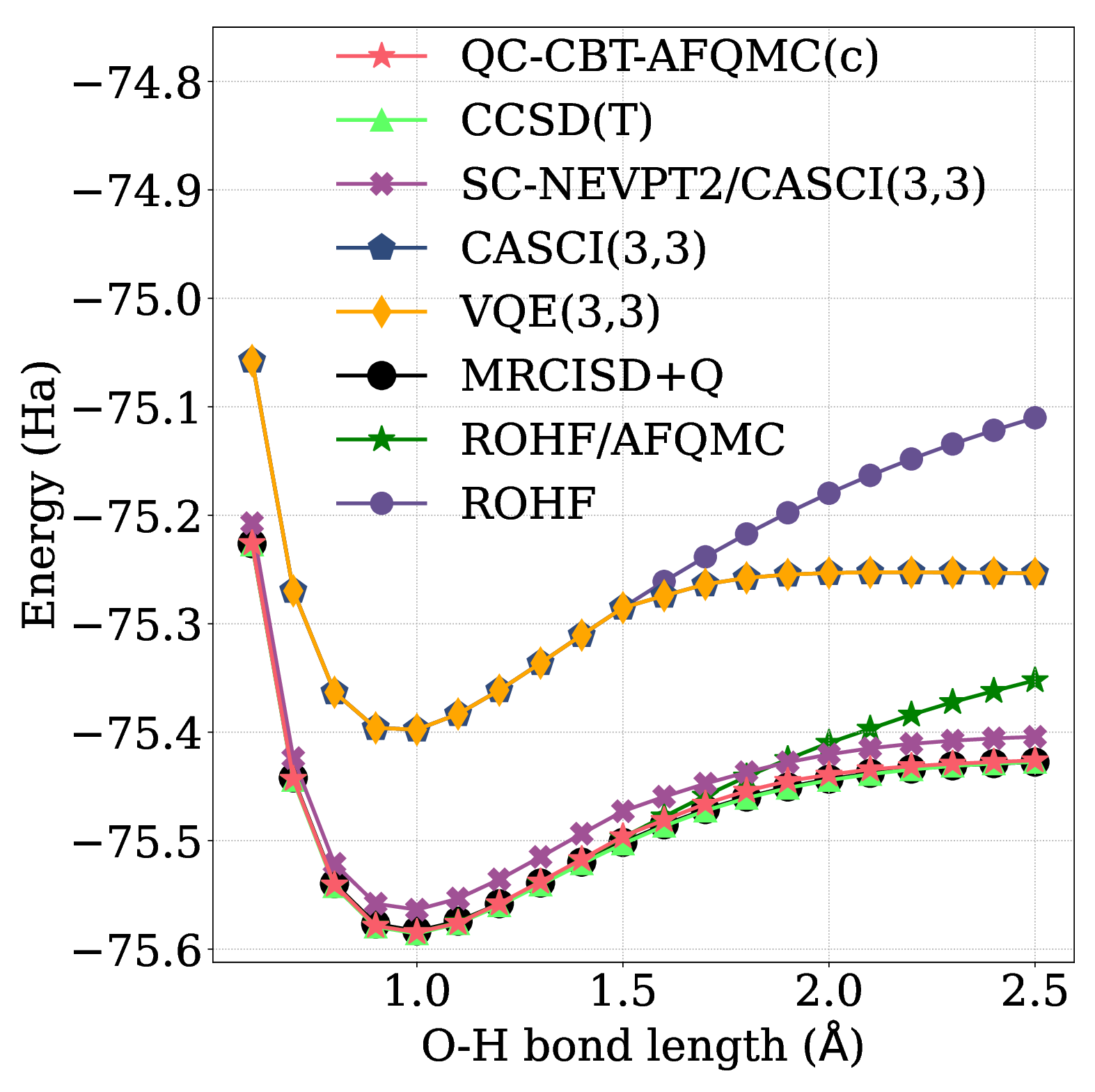}
    \caption{Potential energy curves for the \ch{OH} radical obtained using a ($3$,$3$) active space and the aug-cc-pVDZ basis set. These results highlight how QC-CBT-AFQMC recovers dynamical correlation effects missing from the active space alone, closely matching benchmark methods and accurately describing the bond dissociation.}
\label{OH_PESes}
\end{figure}

\subsection{Potential energy curves}

Initially, the PECs of the hydroxyl radical were mapped as a function of the \ch{O-H} bond length. Utilizing both VQE/UCCSD and CASCI methodologies, we delineated an active space comprising three spatial orbitals to encapsulate the $1$s atomic orbital of the hydrogen and $2$s, and $2$p\textsubscript{x} atomic orbitals of the oxygen that form the $\sigma$-bond, with three electrons shared among them. The selected basis set was the augmented aug-cc-pVDZ Dunning basis, enriched with diffuse functions to accurately capture the formation of radicals during the dissociation process. For the AFQMC calculations, simulations were performed with $10,000$ blocks, where each block corresponds to a fixed number of propagation steps used for statistical sampling and averaging, $10$ steps per block, $480$ walkers, and a time step of $0.005 ~ \text{Ha}^{-1}$. In the CBT phase, we aimed for a maximum of $R = 5$ determinants, setting the measurement counts for $N_f$, $N_a$, and $N_b$ at $10^6$, resulting in a total of approximately $9 \cdot 10^6$ measurements. The maximum number of CNOT gates in the CBT circuits is $4$ for fully connected qubits.

Figure \ref{OH_PESes} presents the constructed PECs for the hydroxyl radical. Notably, the restricted open-shell Hartree--Fock (ROHF) method diverges as the bond length increases indefinitely, highlighting the multireferential nature of this extensively stretched system. A similar behavior is observed with AFQMC based on the ROHF trial wave function. Conversely, both CASCI and its quantum computing counterpart, VQE, accurately capture the energy plateau at long bond lengths, with almost indistinguishable PEC results. However, neither CASCI($3$,$3$) nor VQE($3$,$3$) accurately depicted the PEC shape in the intermediate region (around $1.5 ~ \text{\AA}$), which is due to the absence of dynamic electron correlation. In contrast, the QC-CBT-AFQMC approach yields a PEC similar to that derived from CCSD(T) and MRCISD+Q and matches the SC-NEVPT2/CASCI($3$,$3$) PEC with minimal deviation. Only small discrepancies observed in the intermediate bond length range affirm the reliability of our method, positioning it on par with CCSD(T), MRCISD+Q and NEVPT$2$/CASCI($3$,$3$) for analyzing this system.

Next, PECs were plotted for the ethylene molecule, which was analyzed in terms of the changes in the dihedral \ch{HC-CH} angle, specifically the rotation around the double bond. In the applications of both VQE/UCCSD and CASCI, an active space of $2$ orbitals was defined to encompass the $2$p\textsubscript{z} carbon atomic orbitals that constitute the $\pi$-bond, with $2$ electrons shared between them. The cc-pVDZ Dunning basis set was chosen for this study. For the AFQMC step, the configuration included $10,000$ blocks, $10$ steps per block, $480$ walkers, and a time step of $0.005 ~ \text{Ha}^{-1}$. In the CBT approach, we aimed for a maximum of $R = 2$ determinants, with $N_f$, $N_a$, and $N_b$ set to $10^6$ measurements each, resulting in a total of $3 \cdot 10^6$ measurements. The maximum number of CNOT gates in the CBT circuits is $4$ for fully connected qubits.

\begin{figure}[ht]
    \includegraphics[width=0.45\textwidth]    {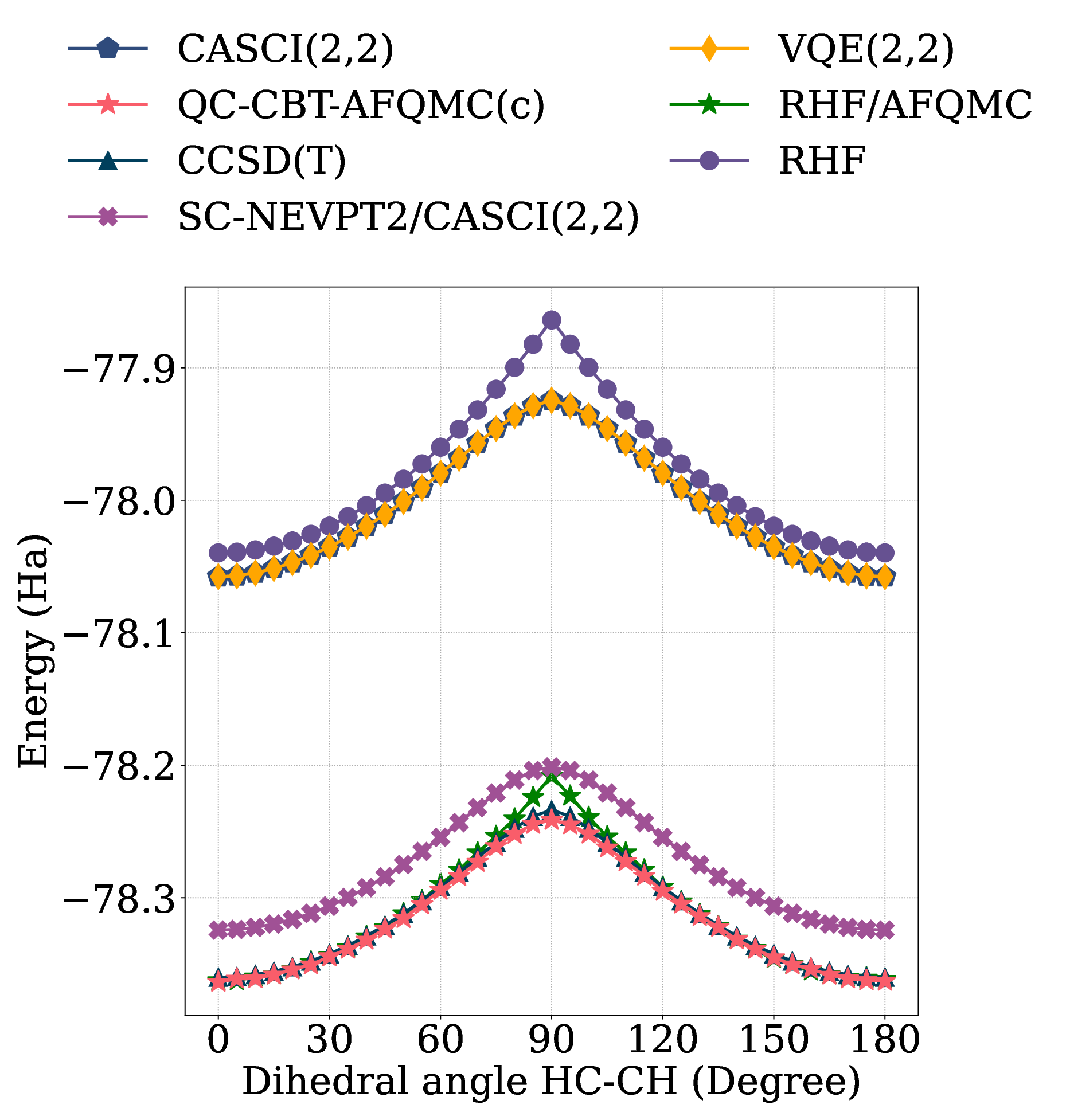}
    \caption{
    Potential energy curves for ethylene as the \ch{HC-CH} dihedral angle varies, computed with a ($2$,$2$) active space and a cc-pVDZ basis set. The QC-CBT-AFQMC(c) approach improves upon CASCI and VQE.
    }
    \label{C2H4_PESes}
\end{figure}

Figure \ref{C2H4_PESes} presents the PECs for the ethylene molecule. The restricted Hartree--Fock (RHF) resulted in a cusp at a dihedral angle of $90^{\circ}$, which is a manifestation of the degeneracy of the energy states. The same incorrect behavior is visible for the AFQMC method, which uses RHF as a trial wave function. In contrast, the CASCI($2$,$2$) and VQE($2$,$2$) methods effectively smooth the energy barrier. However, accurately depicting the overall shape of the PEC necessitates accounting for dynamical electron correlation. MRCISD+Q was not included as a reference in this case, since in a minimal ($2$,$2$) active space it does not adequately describe the dynamical correlation, performing worse than CCSD(T), AFQMC, and even NEVPT$2$. The PEC derived using QC-CBT-AFQMC closely aligns with the CCSD(T) results, except for the region near $90^{\circ}$, where the QC-CBT-AFQMC PEC exhibits a gentler incline. In contrast, the PECs determined using CBT-AFQMC and a downwardly adjusted SC-NEVPT$2$/CASCI($2$,$2$) show almost perfect congruence. The small discrepancy between the CCSD(T) and CBT-AFQMC methods may be attributable to the inherent single reference nature of the CCSD(T) approach.

As a third case study, we explored the dissociation of the \ch{N2} molecule, a scenario characterized by its complex, highly multireferential nature due to the triple bond. For both VQE/UCCSD and CASCI approaches, we selected an active space encompassing $6$ orbitals to include all 2p atomic orbitals forming the \ch{N-N} bond, with $6$ electrons distributed amongst them. The aug-cc-pVDZ Dunning basis set was used because it contains diffuse functions crucial for accurately modeling the dissociation process. The parameters for the AFQMC calculations were set to $10,000$ blocks, $10$ steps per block, $480$ walkers, and a time step of $0.005 ~\text{Ha}^{-1}$. In the CBT phase, we aimed for a maximum of $R = 102$ determinants, with the measurement counts $N_f$, $N_a$, and $N_b$ set at $10^6$ each, culminating in a total of $2.03 \cdot 10^8$ measurements. The maximum number of CNOT gates in the CBT circuits is $10$ for fully connected qubits.

\begin{figure}[ht]
    \includegraphics[width=0.45\textwidth]    {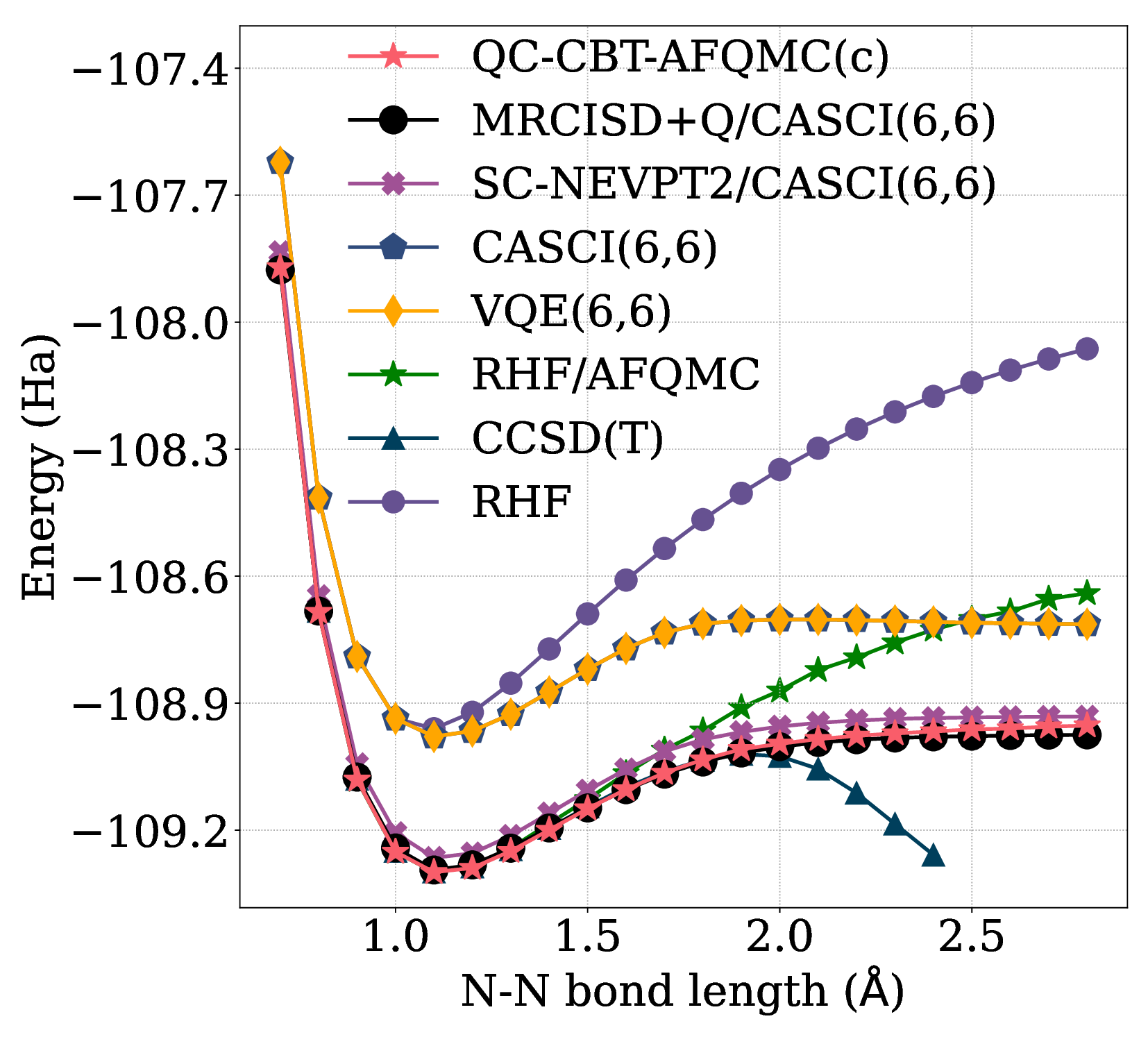}
    \caption{
Potential energy curves for \ch{N2}, obtained with a ($6$,$6$) active space and the aug-cc-pVDZ basis set. QC-CBT-AFQMC(c) accurately captures the multireference character of the stretched \ch{N-N} bond, showing improved agreement with MRCISD+Q references over other methods.
}
    \label{N2_PESes}
\end{figure}

Figure \ref{N2_PESes} displays the PECs constructed for the \ch{N2} molecule. As anticipated, the RHF method cannot accurately predict the PEC for the significantly stretched \ch{N2} molecule. AFQMC also fails when using RHF as a trial wave function. Notably, even the CCSD(T) method exhibits poor performance due to its inherent single-reference limitation and perturbative nature. Consequently, MRCISD+Q was employed as the reference method for this challenging case. Similarly to the OH radical, both CASCI($6$,$6$) and VQE($6$,$6$) accurately predict the energy plateau at extended bond lengths but cannot depict the correct PEC shape in the intermediate region (approximately $1.5 - 2.0 ~ \text{\AA}$). Again, QC-CBT-AFQMC demonstrates commendable agreement with MRCISD+Q across both equilibrium and intermediate regions. Nevertheless, deviations from the expected plateau become apparent at longer distances. This discrepancy is attributed to the fact that while MRCI employs CASSCF($6$,$6$)-optimized orbitals, CBT-AFQMC relies on the same orbitals as VQE($6$,$6$), specifically the RHF canonical orbitals. Achieving closer alignment with MRCI necessitates either expanding the active space or optimizing orbitals in a CASSCF-like fashion. Furthermore, the curvature of the QC-CBT-AFQMC PEC in the intermediate region aligns more closely with MRCI than with SC-NEVPT$2$/CASCI($6$,$6$), highlighting the precision of the proposed method.

Another point worth mentioning is the computation time. As stated in \cite{huang2024evaluatingquantumclassicalquantummonte}, classical postprocessing for the calculation of benzene on a single CPU core would take approximately $1000$ hours per time step using a ($6$,$6$) active space. In our case, the full QC-CBT-AFQMC calculation for a single bond length of \ch{N2} within the same active space took at most around $5.5$ hours on $48$ CPU cores. The CBT part was not parallelized and took only several minutes to complete. This demonstrates the much higher efficiency and practicality of our CBT approach compared to the existing Matchgate shadows method.

\subsection{Effect of measurement quantity on computational basis tomography}

The accuracy of CBT is inherently linked to the number of performed measurements, where more measurements typically yield more precise CI coefficients. However, conducting millions of CBT measurements in practical scenarios is time-consuming. Importantly, the primary objective within the AFQMC step is to capture the dynamical correlation outside of the active space. In this context, it is assumed that errors in CI coefficients mostly affect the correlation outside the active space. By employing equation \ref{cbt-afqmc_1}, we isolated this energy correction by subtracting the AFQMC energy obtained within the active space approximation from that obtained within the full space. This section examines the precision of our method in practical applications.

\begin{figure*}
    \includegraphics[width=1\textwidth]    {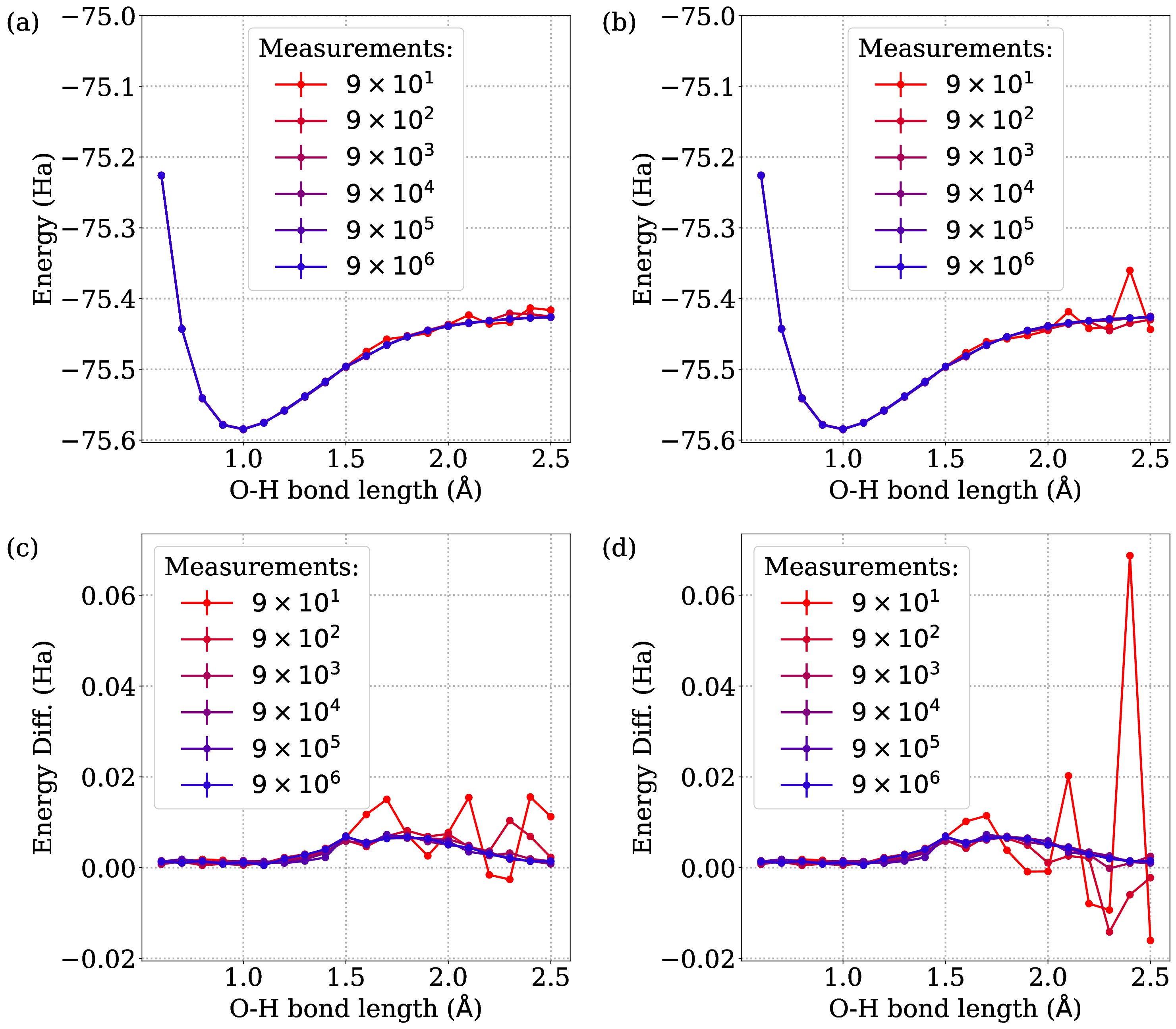}
    \caption{
    Analysis of CBT measurement effects (i.e., statistical errors) on QC-CBT-AFQMC results for the OH radical with a ($3$,$3$) active space and aug-cc-pVDZ basis set. Panels (a) and (b) compare the potential energy curves reconstructed without and with the energy correction from equation \ref{cbt-afqmc_1}, respectively, as the number of CBT measurements changes. Panels (c) and (d) show the corresponding energy deviations from CCSD(T). These results demonstrate that applying the correction significantly reduces measurement-induced errors, allowing accurate computations even with limited shot budgets.
    }
    \label{OH_err}
\end{figure*}

Focusing on the OH radical, the left side of Figure \ref{OH_err} clearly shows the effect of the number of CBT measurements on the AFQMC energy based on the CBT-VQE initial guess. Notably, the error increases with the extension of the \ch{O-H} bond, correlating with the system's increasingly multireferential character. This multireferential aspect should ideally be accounted for during the VQE stage. Consequently, applying equation \ref{cbt-afqmc_1} effectively mitigates these errors, as shown in the left segment of Figure \ref{OH_err}. It is crucial to acknowledge that the PECs obtained using a large number of measurements, whether or not equation \ref{cbt-afqmc_1} is applied, converge closely to the reference because $E^{act}_{AFQMC}$ converges to $E^{act}_{VQE}$.

Other molecular systems were also explored and yielded comparable outcomes. The detailed results of these investigations are available in Appendix \ref{appendix:shots}.

\subsection{Analysis of reaction barriers}

To illustrate the practical applicability of our method, we explored the reaction barriers of [$3$+$2$] cycloaddition reactions as a representative example. These reactions are important in biochemistry, particularly for the \textit{in vivo} study of biomolecules \cite{Jewett2010}. Cycloaddition enables the precise tagging of biomolecules in a process known as bio-orthogonal click reactions. In addition to real-time imaging, bio-orthogonal reactions are indispensable for pioneering applications in medicine and healthcare, such as targeted cancer therapies and \textit{in situ} drug synthesis \cite{devaraj_future_2018}. The search for new, especially mutually orthogonal, reactions is a priority research area.

\begin{figure*}[ht!]
    \makebox[\textwidth][c]{
    \includegraphics[width=0.48\textwidth]{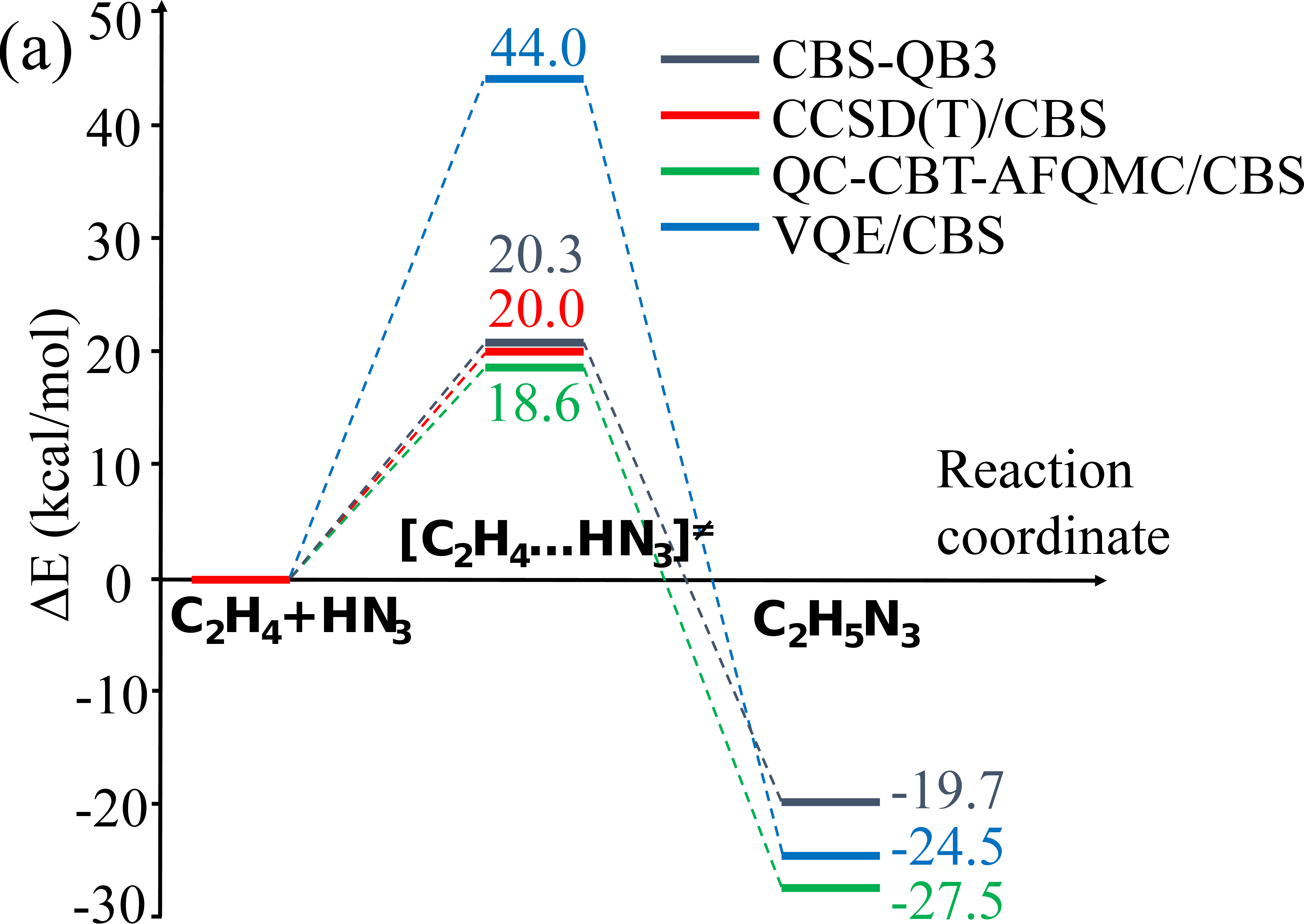}%
    \hspace{0.04\textwidth} 
    \includegraphics[width=0.48\textwidth]{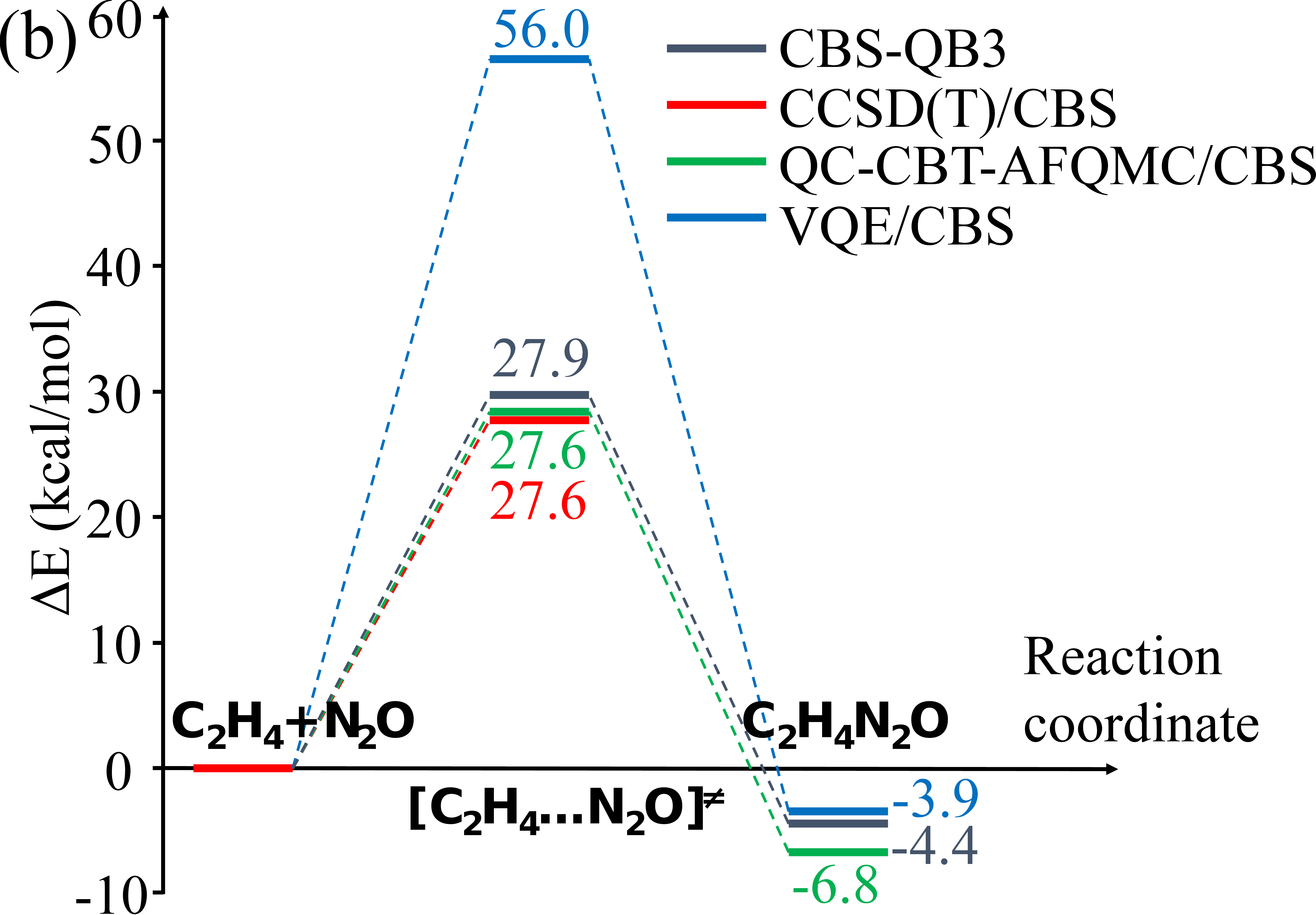}%
}
    \caption{
    Computed energy barriers for the [$3$+$2$] cycloaddition reactions illustrated in Figure \ref{cycloadd_reactions}, comparing QC-CBT-AFQMC results against CCSD(T)/CBS and CBS-QB$3$ reference values. Subfigure (a) shows the results for \ch{HN3}, while subfigure (b) shows those for \ch{N2O}. The close agreement underscores QC-CBT-AFQMC’s ability to accurately estimate reaction barriers and demonstrates the effectiveness of incorporating complete basis set extrapolations.
}
    \label{Reaction_barriers}
\end{figure*}

\begin{figure}[ht!]
    \includegraphics[width=1.00\linewidth]    {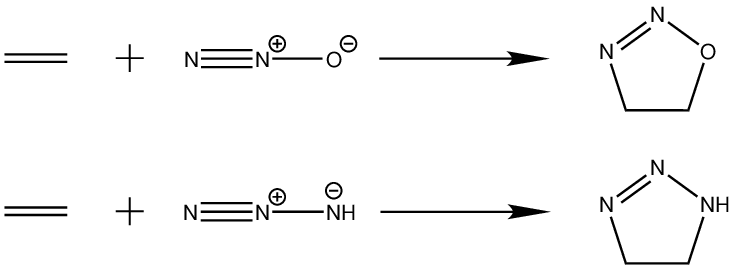}
    \caption{
    Illustration of the [$3$+$2$] cycloaddition reactions studied: ethylene reacting with \ch{N2O} and \ch{HN3}. These reactions serve as test cases for evaluating the accuracy of QC-CBT-AFQMC in computing reaction barriers.}
    \label{cycloadd_reactions}
\end{figure}  

Among the extensive array of reactions, we focused on the relatively simple yet crucial reactions between ethylene and nitrous oxide \ch{N2O} or hydrogen azide \ch{HN3} (Figure \ref{cycloadd_reactions}). The manageable size of these systems permits highly accurate calculations of reaction barriers, enabling the construction of a complete basis set extrapolation. Our findings were benchmarked against the CCSD(T)/CBS \cite{Ess2005} and CBS-QB$3$ \cite{Karton2015} methods, which have been established as reliable approaches for calculating reaction barriers.

To determine the reaction barriers, we conducted single-point calculations for reactants at their equilibrium geometries and for the transition state at the peak of the barrier. All geometric data were derived using density functional theory (DFT) using the B3LYP functional with the def$2$-SVP basis set, as reported in the work by Stuyver et al. \cite{Stuyver2023}. Given the high efficiency of DFT for providing optimal geometries for various high-level theoretical approaches, and considering that the DFT geometries were employed in notable studies \cite{Ess2005, Karton2015}, we adopted a similar strategy.

The initial phase of our investigation involved executing the VQE for each geometric configuration. The VQE wave function was translated exactly, which corresponds to the CBT with an infinite number of measurements. This approximation effectively eliminates any potential errors arising from the measurement process. For the ethylene molecule, we defined an active space with $2$ electrons across $2$ orbitals, encompassing both bonding and anti-bonding $\pi$-orbitals. For \ch{N2O}, the active space included $4$ electrons on $4$ $\pi$-orbitals, with one involved in \ch{N-N} bonding and simultaneously in anti-bonding of \ch{O-H} bonds. This molecule exhibits axial symmetry, leading to sets of $2$ degenerate orbitals, thus requiring the selection of a pair of such orbitals. In the case of \ch{HN3}, an active space with $4$ electrons over $3$ orbitals was chosen, incorporating $2$ $\pi$-bonding and one non-bonding orbital. Given that both the final products and the transition states in these reactions share similar structures, featuring a single double bond, we selected active spaces of $2$ electrons across $2$ orbitals for all entities. These orbitals represent the $\pi$ bonding and anti-bonding interactions associated with the double bond.

In the final phase, we conducted AFQMC calculations, employing VQE-CBT wave functions as the initial guesses. The AFQMC simulations were configured with $10,000$ blocks, $10$ steps per block, $960$ walkers, and a time step of $0.005 ~ \text{Ha}^{-1}$. Moreover, the initial $1,000$ blocks were omitted to allow system equilibration.

To construct a complete basis set extrapolation, we performed all calculations using a sequential series of Dunning basis sets, aug-cc-pVnZ (where n = D, T, Q), enriched with diffuse functions. This series of basis sets is known for its rapid convergence compared to those without diffuse functions. Subsequently, we applied two distinct CBS extrapolation techniques to both the VQE energies and AFQMC correlation energies. Comprehensive details are provided in Appendix \ref{appendix:cbs}. Eventually, we selected an exponential three-basis-set scheme for the VQE energy and the Riemann zeta function method for the AFQMC correlation energy. These results are presented in Figure \ref{Reaction_barriers}.

Figure \ref{Reaction_barriers} shows that the VQE/CBS method generally predicts the thermodynamics of reactions with accuracy approaching that of the more sophisticated perturbation theory-based CBS-QB$3$ method, particularly noticeable in the reaction of \ch{C2H4} with \ch{N2O}. However, VQE struggles to accurately predict the reaction barriers for both reactions. Furthermore, the use of AFQMC gives a more accurate prediction of the reaction barrier energies, even for the smallest basis set. This behavior is expected as reactants and products are single reference closed-shell systems, whereas the transitional state is a highly stretched multireferential system. The Riemann zeta function method yields results comparable to the reference data, similar to the CCSD(T)/CBS for the barrier energy of the \ch{C2H4 + N2O} reaction, albeit with a slight overestimation of the heat effect of the reaction. For the \ch{C2H4 + HN3} reaction, the same approach estimates the barrier energy within $1.5$ kcal/mol of that given by CCSD(T) and the reaction's heat effect within $8$ kcal/mol of that given by CBS-QB$3$. Remarkably, our findings closely align with those obtained using the much more costly and precise CCSDT(Q)/CBS method \cite{Vermeeren2022}. In our analysis of the \ch{C2H4} and \ch{HN3} reaction, we determined an energy barrier of $18.6$ kcal/mol, while the referenced study reported a barrier of $18.29$ kcal/mol for the reaction between \ch{C2H4} and \ch{CH3N3}. This similarity underscores the chemical resemblance between the two systems, with the only possible variance arising from the presence of the \ch{CH3} group, which likely introduces a small geometrical obstacle. Similar findings were observed for the heat effect; our method gave a value of $-27.5$ kcal/mol, which is very close to the $-28.86$ kcal/mol reported in the referenced study, underscoring the high accuracy of our method.

\section{Conclusions}
In this work, we presented QC-CBT-AFQMC, which integrates the quantum-classical auxiliary-field quantum Monte Carlo (QC-AFQMC) approach with computational basis tomography (CBT). Our main motivation stems from the practical limitations of current schemes of quantum tomography under realistic shot budgets. Although QC-AFQMC with Matchgate shadow has polynomial asymptotic scaling, its high polynomial degree ($O(N_q^{8.5} \log^2(N_q))$), circuit requirements, and measurement overhead limit its practicality particularly for near-term quantum devices and could remain non-trivial even in the fault-tolerant quantum computing era. 

In contrast, CBT uses sparse Clifford circuits with $O(N_q)$ CNOT gates and has a lower constant prefactor, allowing us to do calculations for bigger active spaces. CBT circuits can be implemented in constant depth using mid-circuit measurements. The classical postprocessing cost of QC-CBT-AFQMC $O(N_q^4)$ is lower than that of the Matchgate shadows. This makes CBT applicable to near-term quantum devices. In addition, the extracted wave function data via CBT is stored in double precision format, avoiding exponential scaling of the measurements. Within the QC-AFQMC framework, the wave function data measured by CBT are then used to mitigate the sign problem in AFQMC, enabling efficient incorporation of dynamical correlation.

Tests on molecular benchmarks under realistic shot constraints showed that accurate trial wave functions were obtained using CBT. We further applied QC-CBT-AFQMC to [$3$+$2$] cycloaddition reactions, which are important for biochemistry and medicine. The predicted barriers of the method matched those from established methods like CCSD(T)/CBS and CBS-QB$3$. We also examined the performance of various complete basis set (CBS) extrapolation techniques for AFQMC, finding that both the inverse cube and Riemann approaches gave similar and reliable results.

Looking ahead, there are multiple directions for further improvements. Techniques such as QSCI/ADAPT-QSCI~\cite{Kanno2023QSCI,nakagawa2024adapt,robledo2024qsci,barison2024qsci,kaliakin2024qsci,sugisaki2024qsci,mikkelsen2024qsci,yu2025qsci}, which can potentially reduce the circuit depth and the measurement demands, could be integrated to achieve even higher shot efficiency. 
Moreover, additional research is needed to explore the applicability of the QC-CBT-AFQMC method to larger and more complex molecular systems, as well as to investigate the combination of CBT with other quantum-classical hybrid quantum Monte Carlo approaches~\cite{Kanno2023QMC, blunt2024quantumcomputingapproachfixednode,kiser2024contextualsubspaceauxiliaryfieldquantum}.

\begin{acknowledgments}
We thank Yuichiro Yoshida for the fruitful discussions. This project was supported by funding from the MEXT Quantum Leap Flagship Program (MEXTQLEAP) through Grant No. JPMXS0120319794 and the JST COI-NEXT Program through Grant No. JPMJPF2014. The completion of this research was partially facilitated by a JSPS Grants-in-Aid for Scientific Research (KAKENHI), specifically Grant Nos. JP23H03819 and JP21K18933.
We thank the Supercomputer Center, the Institute for Solid State Physics of the University of Tokyo for the use of their facilities. 
This work was also achieved through the use of the SQUID system at the Cybermedia Center, the University of Osaka.
\end{acknowledgments}

\appendix

\section{Dependency analysis of QC-CBT-AFQMC on the number of measurements for ethylene and \ch{N2}}
\label{appendix:shots}

\begin{figure*}
    \includegraphics[width=1\textwidth]    {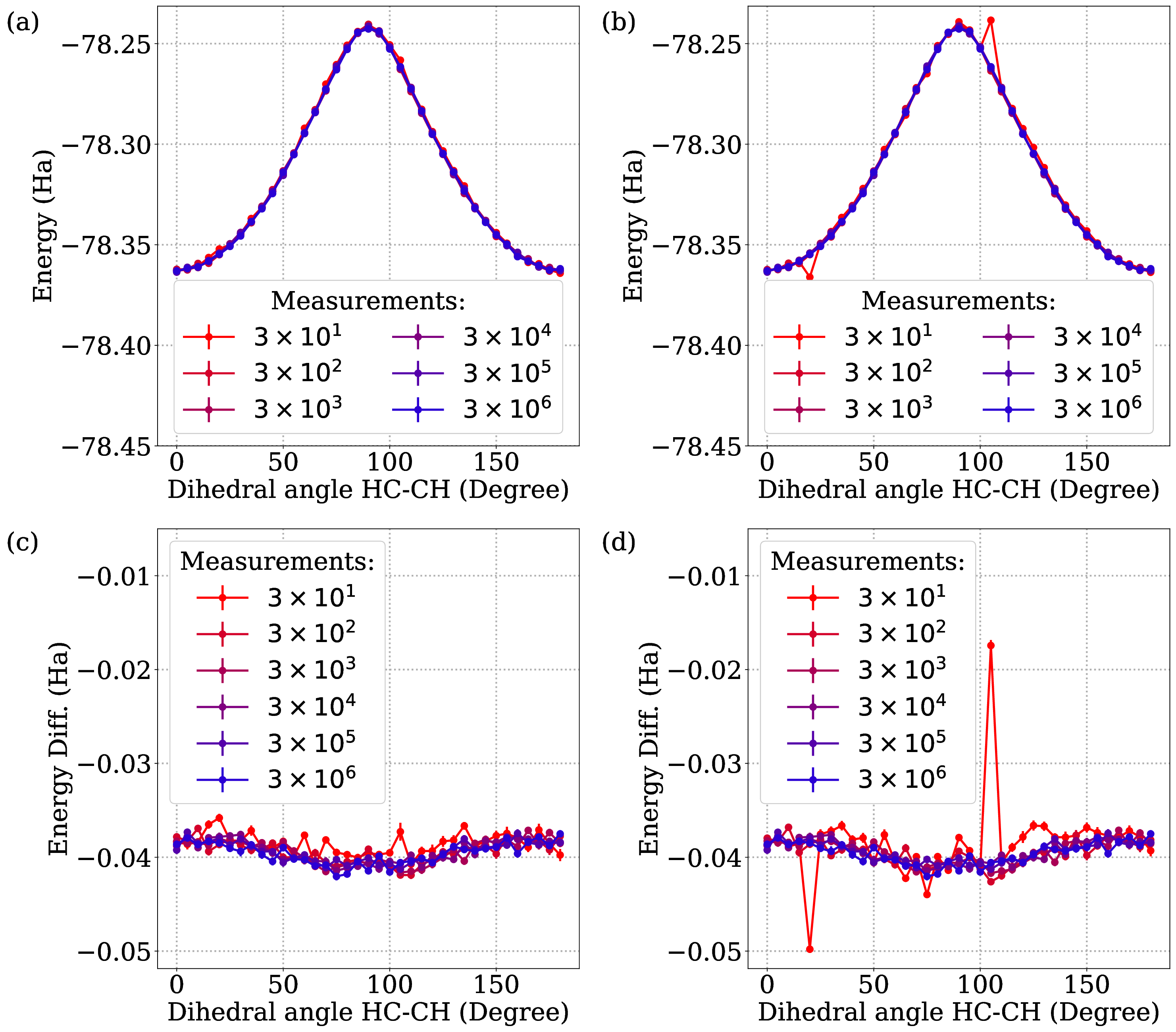}
    \caption{Potential energy curves for the ethylene molecule in the ($2$,$2$) active space with the cc-pVDZ basis set as a function of the number of CBT measurements (a) without and (b) with applying equation \ref{cbt-afqmc_1}. Energy differences between QC-CBT-AFQMC and SC-NEVPT$2$/CASCI($2$,$2$) as a function of the number of CBT measurements (c) without and (d) with applying equation \ref{cbt-afqmc_1}.}
    \label{C2H4_err}
\end{figure*}

For ethylene, we benchmarked our CBT-AFQMC results against SC-NEVPT$2$ and CCSD(T) methods (refer to the lower part of Figure \ref{C2H4_err}). In the SC-NEVPT$2$ analysis, the energy does not vary significantly at the $90^{\circ}$ dihedral angle compared to CCSD(T), albeit with an offset in the energy scale. Nevertheless, both analyses reveal minimal energy discrepancies. This was expected since our system is predominantly a single reference across most geometries. However, the presence of two anomalies was corrected using the formula \ref{cbt-afqmc_1}.

\begin{figure*}
    \includegraphics[width=1\textwidth]    {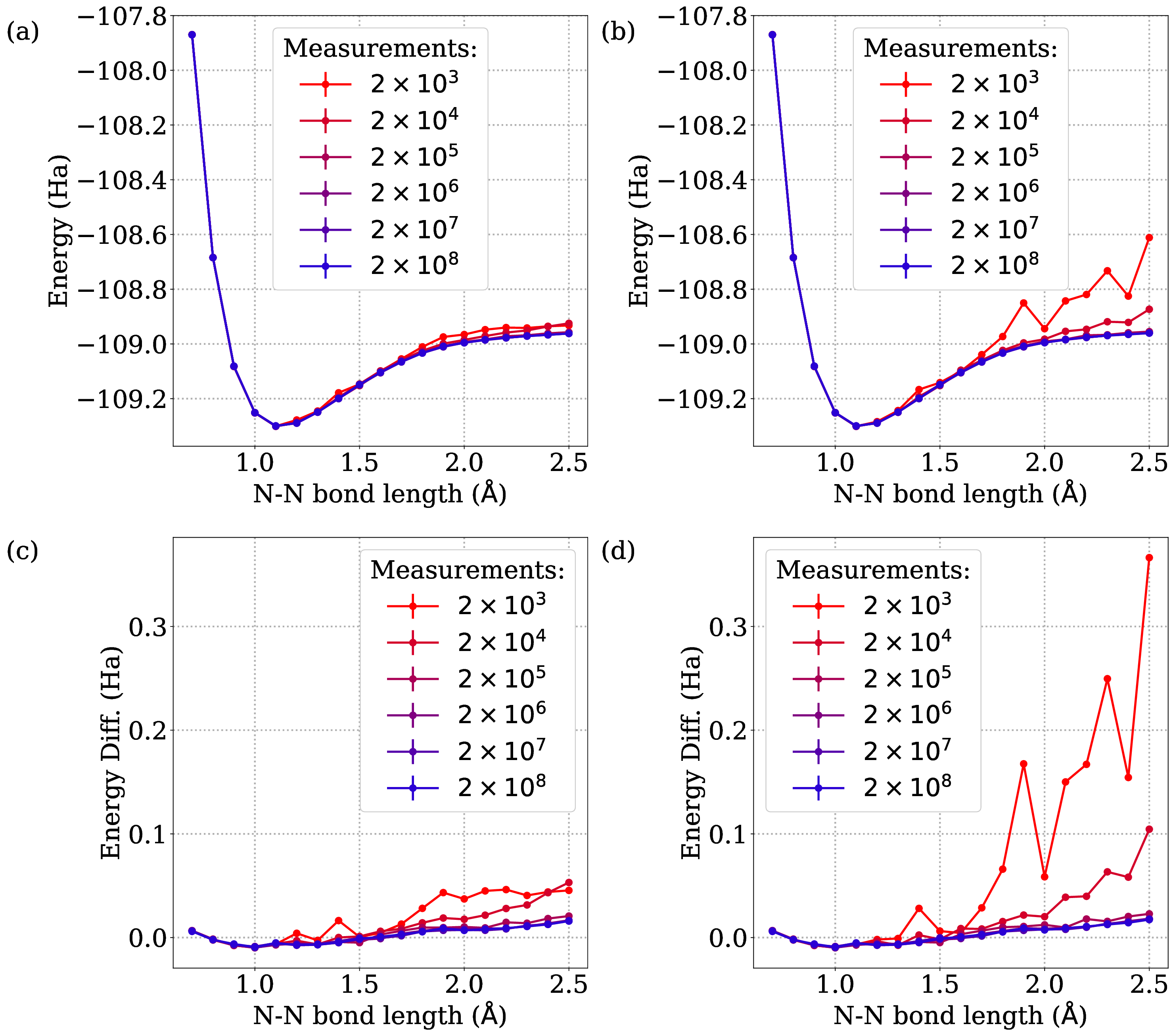}
    \caption{Potential energy curves for the \ch{N2} molecule in the ($6$,$6$) active space with the aug-cc-pVDZ basis set as a function of the number of CBT measurements (a) without and (b) with implementing equation \ref{cbt-afqmc_1}. Energy differences between QC-CBT-AFQMC and MR-CISD+Q as a function of the number of CBT measurements (c) without and (d) with using equation \ref{cbt-afqmc_1}.}
    \label{N2_err}
\end{figure*}

As previously highlighted, the nitrogen molecule represents a challenging multi-reference scenario at extended \ch{N-N} bond lengths. Figure \ref{N2_err} illustrates the significant escalation of errors as the bond length expands. Nonetheless, by implementing equation \ref{cbt-afqmc_1}, we successfully mitigated these errors, even with a small number of measurements. This reduction is feasible because the VQE component effectively captures most of the static correlation.

\section{Complete basis set extrapolation}
\label{appendix:cbs}

The incompleteness of the basis set significantly contributes to discrepancies between experimental results and computational calculations. Therefore, CBS extrapolation is crucial for achieving accurate and practically viable results. Additionally, reference calculations were conducted with CBS extrapolation, prompting us to adopt a similar strategy.

Typically, for post-Hartree--Fock calculations, it is advisable to differentiate between the SCF energy ($E^{SCF}_X$) and the correlation energy ($E^{corr}_X$):
\begin{equation} \label{cbs_1}
    E^{tot}_X = E^{SCF}_X + E^{corr}_X,  
\end{equation}
where $E^{tot}_X$ denotes the total energy of the method under consideration, and $X$ indicates the cardinal number of the basis set. Subsequently, distinct CBS schemes are applied to each component of the total energy because the convergence rate of $E^{SCF}_X$ is typically much faster than that of $E^{corr}_X$. In our study, $E^{SCF}_X$ corresponds to the VQE energy, analogous to CASCI, while $E^{corr}_X$ represents the energy difference between the energies derived using AFQMC and VQE, signifying the AFQMC correlation energy. For the convergence analysis of VQE energy, we employed SCF extrapolation schemes. Table \ref{CBS_table_2} illustrates the rapid convergence of VQE energies with increasing basis set size, showing a pattern similar to that observed with SCF convergence.

Several existing techniques for CBS extrapolation are applicable to SCF and CASSCF. Initially, we employed the exponential scheme \cite{halkier_basis-set_1999} as described by: 
\begin{equation} \label{cbs_2}
    E^{SCF}_X = E^{SCF}_{\infty} + A \exp(-\beta X),  
\end{equation}
where $\beta$, $A$, and $E^{SCF}_{\infty}$ are parameters determined through fitting. This method utilizes data from all three basis sets. The second approach adapts the exponential formula, incorporating pre-optimized parameters tailored for the CASSCF method based on molecular datasets \cite{Pansini2016}. The formula for this technique is given by:
\begin{equation} \label{cbs_3}
    E^{CAS}_{\infty} = \frac{E_{X_i} e^{\beta X_i} - E_{X_j} e^{\beta X_j}}{e^{\beta X_i} - e^{\beta X_j}}, 
\end{equation}
where $X_i$ represents the optimized cardinal numbers of the basis sets, which slightly deviate from their integer values, and $\beta$ is another parameter obtained through fitting. For our extrapolations, we chose the aug-cc-pVTZ and aug-cc-pVQZ basis sets. Despite the different approaches, both methods yielded similar results, with a marginal discrepancy of only $1$ mHa, as detailed in Table \ref{CBS_table_1}.

\begin{table*}[ht]
\caption{Calculated energies for all systems used in reaction barrier calculations (Hartree). For AFQMC calculations, only correlation energies are provided (Hartree). exp($2$,$3$,$4$) is method \ref{cbs_2}, SCF-E($3$,$4$) is method \ref{cbs_3}, $\zeta$($2$,$3$,$4$) is method \ref{cbs_4} and X\textsuperscript{$-3$}($3$,$4$) is method \ref{cbs_6}. nZ refers to the aug-cc-pVnZ basis set.}
\label{CBS_table_1}
\resizebox{\textwidth}{!}{
\begin{tabularx}{\textwidth}{XXXXXXXX}
\hline
\hline
Method/ Basis           & C\textsubscript{$2$}H\textsubscript{$4$}     & N\textsubscript{$2$}O       & HN\textsubscript{$3$}       & TS-C\textsubscript{$2$}H\textsubscript{$4$}-N\textsubscript{$2$}O & TS-C\textsubscript{$2$}H\textsubscript{$4$}-HN\textsubscript{$3$} & C\textsubscript{$2$}H\textsubscript{$4$}N\textsubscript{$2$}O   & C\textsubscript{$2$}H\textsubscript{$5$}N\textsubscript{$3$}    \\
\hline
\hline
CASCI/ DZ         & $-78.0431$ & $-183.7117$ & $-163.8674$ & $-261.6730$   & $-241.8448$   & $-261.7693$ & $-241.9556$ \\
CASCI/ TZ         & $-78.0637$ & $-183.7558$ & $-163.9041$ & $-261.7311$   & $-241.8981$   & $-261.8267$ & $-242.0066$ \\
CASCI/ QZ         & $-78.0683$ & $-183.7679$ & $-163.9143$ & $-261.7470$   & $-241.9124$   & $-261.8425$ & $-242.0211$ \\
\hline
CASCI/ exp($2$,$3$,$4$) & $-78.0697$ & $-183.7725$ & $-163.9181$ & $-261.7529$   & $-241.9176$   & $-261.848$5 & $-242.0268$ \\
CASCI/ SCF-E($3$,$4$) & $-78.0697$ & $-183.7715$ & $-163.9173$ & $-261.7517$   & $-241.9166$   & $-261.8472$ & $-242.0254$ \\
\hline
AFQMC/ DZ         & $-0.3333$  & $-0.5649$   & $-0.5589$   & $-0.9485$     & $-0.9322$     & $-0.9062$   & $-0.8900$   \\
AFQMC/ TZ         & $-0.4186$  & $-0.7156$   & $-0.7012$   & $-1.1861$     & $-1.1630$     & $-1.1459$   & $-1.1258$   \\
AFQMC/ QZ         & $-0.4562$  & $-0.7955$   & $-0.7740$   & $-1.3011$     & $-1.2726$     & $-1.2609$   & $-1.2363 $  \\
\hline
AFQMC/ $\zeta$($2$,$3$,$4$)          & $-0.4974$  & $-0.8889$   & $-0.8582$   & $-1.4315$     & $-1.3961$     & $-1.3908$   & $-1.3603$   \\
AFQMC/ X\textsuperscript{$-3$}($3$,$4$)    & $-0.4837$  & $-0.8537$   & $-0.8271$   & $-1.3849$     & $-1.3526$     & $-1.3447$   & $-1.3170$ \\
\hline
\hline
\end{tabularx}
}
\end{table*}

\begin{table*}[ht]
\caption{Reaction barrier energies in different basis sets and under different CBS extrapolations (kcal/mol). exp($2$,$3$,$4$) is method \ref{cbs_2}, SCF-E($3$,$4$) is method \ref{cbs_3}, $\zeta$($2$,$3$,$4$) is method \ref{cbs_4} and X\textsuperscript{$-3$}($3$,$4$) is method \ref{cbs_6}. nZ refers to the aug-cc-pVnZ basis set.}
\label{CBS_table_2}
\begin{tabularx}{\textwidth}{XXXXX}
\hline
\hline
Method/Basis                                      & C\textsubscript{$2$}H\textsubscript{$4$}$+$N\textsubscript{$2$}O Barrier Energy & C\textsubscript{$2$}H\textsubscript{$4$}$+$N\textsubscript{$2$}O Reaction Energy & C\textsubscript{$2$}H\textsubscript{$4$}$+$HN\textsubscript{$3$} Barrier Energy & C\textsubscript{$2$}H\textsubscript{$4$}$+$HN\textsubscript{$3$} Reaction Energy \\
\hline
\hline
CASCI/DZ                                    & $51.4$                                & $-9.1$                                 & $41.3$                                & $-28.2$                                \\
CASCI/TZ                                    & $55.4$                                & $-4.6$                                 & $43.8$                                & $-24.3$                                \\
CASCI/QZ                                    & $56.0$                                & $-3.9$                                 & $44.1$                                & $-24.1$                                \\
\hline
CASCI/exp($2$,$3$,$4$)                            & $56.0$                                & $-3.9$                                 & $44.0$                                & $-24.5$                                \\
CASCI/SCF-E($3$,$4$)                            & $56.2$                                & $-3.7$                                 & $44.2$                                & $-24.1$                                \\
\hline
AFQMC/DZ                                    & $19.8$                                & $-14.1$                                & $16.2$                                & $-26.8$                                \\
AFQMC/TZ                                    & $22.9$                                & $-11.9$                                & $16.7$                                & $-28.0$                                \\
AFQMC/QZ                                    & $25.0$                                & $-9.7$                                 & $17.5$                                & $-28.0$                                \\
\hline
CASCI/exp($2$,$3$,$4$) $+$ AFQMC/$\zeta$($2$,$3$,$4$)              & $27.6$                                & $-6.8$                                 & $18.6$                                & $-27.5$                                \\
CASCI/SCF-E($3$,$4$) $+$ AFQMC/X\textsuperscript{$-3$}($3$,$4$) & $26.3$                                & $-8.4$                                 & $17.9$                                & $-28.0$                                \\
\hline
CCSD(T)/CBS \cite{Ess2005}                                 & $27.6$                                & No data                                    & $20.0$                                & No data                                    \\
CBS-QB$3$ \cite{Karton2015}                                  & $27.9$                                & $-4.4$                                 & $20.3$                                & $-19.7$                               \\
\hline
\hline
\end{tabularx}
\end{table*}

Furthermore, various methods are available for extrapolating the correlation energy to the CBS limit. Among these, we explored the use of a recently introduced CBS scheme that utilizes the Riemann zeta function, a method known for its generality and solid theoretical foundation \cite{Lesiuk2019}. The principal formula for extrapolating across three consecutive basis sets using this method is expressed as:
\begin{equation} \label{cbs_4}
    E_{\infty} = E_L + a(\zeta(4) - \sum^L_{l=1} l^{-4}) + b(\zeta(6) - \sum^L_{l=1} l^{-6}),  
\end{equation}
where $\zeta(x)$ denotes the Riemann zeta function (notably, $\zeta(4) = \frac{\pi^4}{90}$ and $\zeta(6) = \frac{\pi^6}{945}$). The coefficients $a$ and $b$ are determined as follows:
\begin{equation} \label{cbs_5}
    \begin{split}
    &a = \frac{L^6(E_L - E_{L-1}) - (L-1)^6(E_{L-1} - E_{L-2})}{2L-1},
    \\
    &b = L^6(E_L - E_{L-1}) - aL^2.
    \end{split} 
\end{equation}
Here, $L$ represents the maximum angular momentum achievable with the chosen basis set. For Dunning's basis sets, $L$ corresponds to the cardinal number of the set. For example, in the aug-cc-pVnZ or cc-pVnZ series, $L=n$.

In addition to the Riemann zeta function approach, we applied the well-established CBS extrapolation method proposed by Helgaker at al. \cite{Helgaker1997}, which uses an inverse cubic function with two fitting parameters:
\begin{equation} \label{cbs_6}
    E_X = E_{\infty} + \frac{a}{X^3},  
\end{equation}
where $X$ is the cardinal number of the basis set. Since this extrapolation formula necessitates the use of two basis sets, aug-cc-pVTZ and aug-cc-pVQZ were used. This scheme has been previously employed in studies focusing on the AFQMC method and has yielded plausible results \cite{Amsler2023}.

Table \ref{CBS_table_1} shows that the CASCI energies exhibit rapid convergence as the basis set size increases. The CBS methods, initially designed for CASSCF but applied to CASCI in our study, yield remarkably consistent energies, diverging by no more than $1.5$ mHa. Such consistency underscores the reliability of the obtained results.

Conversely, the convergence of AFQMC correlation energies is notably slower. The CBS methods are tailored for correlation energy calculations, resulting in more substantial discrepancies in energies across different molecular systems. Nonetheless, when these methods are applied to computing energy barriers, the differences tend to offset each other, resulting in energy discrepancies within $1.5$ kcal/mol, as detailed in Table \ref{CBS_table_2}.

\bibliography{main.bib}

\begin{thebibliography}{60}%
\makeatletter
\providecommand \@ifxundefined [1]{%
 \@ifx{#1\undefined}
}%
\providecommand \@ifnum [1]{%
 \ifnum #1\expandafter \@firstoftwo
 \else \expandafter \@secondoftwo
 \fi
}%
\providecommand \@ifx [1]{%
 \ifx #1\expandafter \@firstoftwo
 \else \expandafter \@secondoftwo
 \fi
}%
\providecommand \natexlab [1]{#1}%
\providecommand \enquote  [1]{``#1''}%
\providecommand \bibnamefont  [1]{#1}%
\providecommand \bibfnamefont [1]{#1}%
\providecommand \citenamefont [1]{#1}%
\providecommand \href@noop [0]{\@secondoftwo}%
\providecommand \href [0]{\begingroup \@sanitize@url \@href}%
\providecommand \@href[1]{\@@startlink{#1}\@@href}%
\providecommand \@@href[1]{\endgroup#1\@@endlink}%
\providecommand \@sanitize@url [0]{\catcode `\\12\catcode `\$12\catcode `\&12\catcode `\#12\catcode `\^12\catcode `\_12\catcode `\%12\relax}%
\providecommand \@@startlink[1]{}%
\providecommand \@@endlink[0]{}%
\providecommand \url  [0]{\begingroup\@sanitize@url \@url }%
\providecommand \@url [1]{\endgroup\@href {#1}{\urlprefix }}%
\providecommand \urlprefix  [0]{URL }%
\providecommand \Eprint [0]{\href }%
\providecommand \doibase [0]{https://doi.org/}%
\providecommand \selectlanguage [0]{\@gobble}%
\providecommand \bibinfo  [0]{\@secondoftwo}%
\providecommand \bibfield  [0]{\@secondoftwo}%
\providecommand \translation [1]{[#1]}%
\providecommand \BibitemOpen [0]{}%
\providecommand \bibitemStop [0]{}%
\providecommand \bibitemNoStop [0]{.\EOS\space}%
\providecommand \EOS [0]{\spacefactor3000\relax}%
\providecommand \BibitemShut  [1]{\csname bibitem#1\endcsname}%
\let\auto@bib@innerbib\@empty
\bibitem [{\citenamefont {Bauer}\ \emph {et~al.}(2020)\citenamefont {Bauer}, \citenamefont {Bravyi}, \citenamefont {Motta},\ and\ \citenamefont {Chan}}]{Bauer2020}%
  \BibitemOpen
  \bibfield  {author} {\bibinfo {author} {\bibfnamefont {B.}~\bibnamefont {Bauer}}, \bibinfo {author} {\bibfnamefont {S.}~\bibnamefont {Bravyi}}, \bibinfo {author} {\bibfnamefont {M.}~\bibnamefont {Motta}},\ and\ \bibinfo {author} {\bibfnamefont {G.~K.-L.}\ \bibnamefont {Chan}},\ }\bibfield  {title} {\bibinfo {title} {Quantum algorithms for quantum chemistry and quantum materials science},\ }\href {https://doi.org/10.1021/acs.chemrev.9b00829} {\bibfield  {journal} {\bibinfo  {journal} {Chem. Rev.}\ }\textbf {\bibinfo {volume} {120}},\ \bibinfo {pages} {12685} (\bibinfo {year} {2020})}\BibitemShut {NoStop}%
\bibitem [{\citenamefont {Motta}\ and\ \citenamefont {Rice}(2022)}]{Motta_2021}%
  \BibitemOpen
  \bibfield  {author} {\bibinfo {author} {\bibfnamefont {M.}~\bibnamefont {Motta}}\ and\ \bibinfo {author} {\bibfnamefont {J.~E.}\ \bibnamefont {Rice}},\ }\bibfield  {title} {\bibinfo {title} {Emerging quantum computing algorithms for quantum chemistry},\ }\href {https://doi.org/https://doi.org/10.1002/wcms.1580} {\bibfield  {journal} {\bibinfo  {journal} {WIREs Comput. Mol. Sci.}\ }\textbf {\bibinfo {volume} {12}},\ \bibinfo {pages} {e1580} (\bibinfo {year} {2022})}\BibitemShut {NoStop}%
\bibitem [{\citenamefont {Peruzzo}\ \emph {et~al.}(2014)\citenamefont {Peruzzo}, \citenamefont {McClean}, \citenamefont {Shadbolt}, \citenamefont {Yung}, \citenamefont {Zhou}, \citenamefont {Love}, \citenamefont {Aspuru-Guzik},\ and\ \citenamefont {O’Brien}}]{Peruzzo2014}%
  \BibitemOpen
  \bibfield  {author} {\bibinfo {author} {\bibfnamefont {A.}~\bibnamefont {Peruzzo}}, \bibinfo {author} {\bibfnamefont {J.}~\bibnamefont {McClean}}, \bibinfo {author} {\bibfnamefont {P.}~\bibnamefont {Shadbolt}}, \bibinfo {author} {\bibfnamefont {M.-H.}\ \bibnamefont {Yung}}, \bibinfo {author} {\bibfnamefont {X.-Q.}\ \bibnamefont {Zhou}}, \bibinfo {author} {\bibfnamefont {P.~J.}\ \bibnamefont {Love}}, \bibinfo {author} {\bibfnamefont {A.}~\bibnamefont {Aspuru-Guzik}},\ and\ \bibinfo {author} {\bibfnamefont {J.~L.}\ \bibnamefont {O’Brien}},\ }\bibfield  {title} {\bibinfo {title} {A variational eigenvalue solver on a photonic quantum processor},\ }\href {https://doi.org/10.1038/ncomms5213} {\bibfield  {journal} {\bibinfo  {journal} {Nat. Commun.}\ }\textbf {\bibinfo {volume} {5}},\ \bibinfo {pages} {4213} (\bibinfo {year} {2014})}\BibitemShut {NoStop}%
\bibitem [{\citenamefont {Lloyd}(1996)}]{Lloyd1996}%
  \BibitemOpen
  \bibfield  {author} {\bibinfo {author} {\bibfnamefont {S.}~\bibnamefont {Lloyd}},\ }\bibfield  {title} {\bibinfo {title} {Universal quantum simulators},\ }\href {https://doi.org/10.1126/science.273.5278.1073} {\bibfield  {journal} {\bibinfo  {journal} {Science}\ }\textbf {\bibinfo {volume} {273}},\ \bibinfo {pages} {1073} (\bibinfo {year} {1996})}\BibitemShut {NoStop}%
\bibitem [{\citenamefont {Aspuru-Guzik}\ \emph {et~al.}(2005)\citenamefont {Aspuru-Guzik}, \citenamefont {Dutoi}, \citenamefont {Love},\ and\ \citenamefont {Head-Gordon}}]{Aspuru2005}%
  \BibitemOpen
  \bibfield  {author} {\bibinfo {author} {\bibfnamefont {A.}~\bibnamefont {Aspuru-Guzik}}, \bibinfo {author} {\bibfnamefont {A.~D.}\ \bibnamefont {Dutoi}}, \bibinfo {author} {\bibfnamefont {P.~J.}\ \bibnamefont {Love}},\ and\ \bibinfo {author} {\bibfnamefont {M.}~\bibnamefont {Head-Gordon}},\ }\bibfield  {title} {\bibinfo {title} {Simulated quantum computation of molecular energies},\ }\href {https://doi.org/10.1126/science.1113479} {\bibfield  {journal} {\bibinfo  {journal} {Science}\ }\textbf {\bibinfo {volume} {309}},\ \bibinfo {pages} {1704} (\bibinfo {year} {2005})}\BibitemShut {NoStop}%
\bibitem [{\citenamefont {Krompiec}\ and\ \citenamefont {Ramo}(2022)}]{Krompiec2022}%
  \BibitemOpen
  \bibfield  {author} {\bibinfo {author} {\bibfnamefont {M.}~\bibnamefont {Krompiec}}\ and\ \bibinfo {author} {\bibfnamefont {D.~M.}\ \bibnamefont {Ramo}},\ }\bibfield  {title} {\bibinfo {title} {Strongly contracted n-electron valence state perturbation theory using reduced density matrices from a quantum computer},\ }\href {https://arxiv.org/abs/2210.05702} {\bibfield  {journal} {\bibinfo  {journal} {arXiv preprint arXiv:2210.05702}\ } (\bibinfo {year} {2022})}\BibitemShut {NoStop}%
\bibitem [{\citenamefont {Tammaro}\ \emph {et~al.}(2023)\citenamefont {Tammaro}, \citenamefont {Galli}, \citenamefont {Rice},\ and\ \citenamefont {Motta}}]{Tammaro2023}%
  \BibitemOpen
  \bibfield  {author} {\bibinfo {author} {\bibfnamefont {A.}~\bibnamefont {Tammaro}}, \bibinfo {author} {\bibfnamefont {D.~E.}\ \bibnamefont {Galli}}, \bibinfo {author} {\bibfnamefont {J.~E.}\ \bibnamefont {Rice}},\ and\ \bibinfo {author} {\bibfnamefont {M.}~\bibnamefont {Motta}},\ }\bibfield  {title} {\bibinfo {title} {-electron valence perturbation theory with reference wave functions from quantum computing: Application to the relative stability of hydroxide anion and hydroxyl radical},\ }\href {https://doi.org/10.1021/acs.jpca.2c07653} {\bibfield  {journal} {\bibinfo  {journal} {J. Phys. Chem. A}\ }\textbf {\bibinfo {volume} {127}},\ \bibinfo {pages} {817} (\bibinfo {year} {2023})}\BibitemShut {NoStop}%
\bibitem [{\citenamefont {Mitarai}\ \emph {et~al.}(2023)\citenamefont {Mitarai}, \citenamefont {Toyoizumi},\ and\ \citenamefont {Mizukami}}]{Mitarai2023}%
  \BibitemOpen
  \bibfield  {author} {\bibinfo {author} {\bibfnamefont {K.}~\bibnamefont {Mitarai}}, \bibinfo {author} {\bibfnamefont {K.}~\bibnamefont {Toyoizumi}},\ and\ \bibinfo {author} {\bibfnamefont {W.}~\bibnamefont {Mizukami}},\ }\bibfield  {title} {\bibinfo {title} {Perturbation theory with quantum signal processing},\ }\href {https://doi.org/10.22331/q-2023-05-12-1000} {\bibfield  {journal} {\bibinfo  {journal} {Quantum}\ }\textbf {\bibinfo {volume} {7}},\ \bibinfo {pages} {1000} (\bibinfo {year} {2023})}\BibitemShut {NoStop}%
\bibitem [{\citenamefont {Cortes}\ \emph {et~al.}(2024)\citenamefont {Cortes}, \citenamefont {Loipersberger}, \citenamefont {Parrish}, \citenamefont {Morley-Short}, \citenamefont {Pol}, \citenamefont {Sim}, \citenamefont {Steudtner}, \citenamefont {Tautermann}, \citenamefont {Degroote}, \citenamefont {Moll}, \citenamefont {Santagati},\ and\ \citenamefont {Streif}}]{Cortes2023}%
  \BibitemOpen
  \bibfield  {author} {\bibinfo {author} {\bibfnamefont {C.~L.}\ \bibnamefont {Cortes}}, \bibinfo {author} {\bibfnamefont {M.}~\bibnamefont {Loipersberger}}, \bibinfo {author} {\bibfnamefont {R.~M.}\ \bibnamefont {Parrish}}, \bibinfo {author} {\bibfnamefont {S.}~\bibnamefont {Morley-Short}}, \bibinfo {author} {\bibfnamefont {W.}~\bibnamefont {Pol}}, \bibinfo {author} {\bibfnamefont {S.}~\bibnamefont {Sim}}, \bibinfo {author} {\bibfnamefont {M.}~\bibnamefont {Steudtner}}, \bibinfo {author} {\bibfnamefont {C.~S.}\ \bibnamefont {Tautermann}}, \bibinfo {author} {\bibfnamefont {M.}~\bibnamefont {Degroote}}, \bibinfo {author} {\bibfnamefont {N.}~\bibnamefont {Moll}}, \bibinfo {author} {\bibfnamefont {R.}~\bibnamefont {Santagati}},\ and\ \bibinfo {author} {\bibfnamefont {M.}~\bibnamefont {Streif}},\ }\bibfield  {title} {\bibinfo {title} {Fault-tolerant quantum algorithm for symmetry-adapted perturbation theory},\ }\href {https://doi.org/10.1103/PRXQuantum.5.010336} {\bibfield  {journal} {\bibinfo  {journal} {PRX
  Quantum}\ }\textbf {\bibinfo {volume} {5}},\ \bibinfo {pages} {010336} (\bibinfo {year} {2024})}\BibitemShut {NoStop}%
\bibitem [{\citenamefont {Nishio}\ \emph {et~al.}(2023)\citenamefont {Nishio}, \citenamefont {Oba},\ and\ \citenamefont {Kurashige}}]{Nishio2023}%
  \BibitemOpen
  \bibfield  {author} {\bibinfo {author} {\bibfnamefont {S.}~\bibnamefont {Nishio}}, \bibinfo {author} {\bibfnamefont {Y.}~\bibnamefont {Oba}},\ and\ \bibinfo {author} {\bibfnamefont {Y.}~\bibnamefont {Kurashige}},\ }\bibfield  {title} {\bibinfo {title} {Statistical errors in reduced density matrices sampled from quantum circuit simulation and the impact on multireference perturbation theory},\ }\href {https://doi.org/10.1039/D3CP03520D} {\bibfield  {journal} {\bibinfo  {journal} {Phys. Chem. Chem. Phys.}\ }\textbf {\bibinfo {volume} {25}},\ \bibinfo {pages} {30525} (\bibinfo {year} {2023})}\BibitemShut {NoStop}%
\bibitem [{\citenamefont {Schleich}\ \emph {et~al.}(2022)\citenamefont {Schleich}, \citenamefont {Kottmann},\ and\ \citenamefont {Aspuru-Guzik}}]{Schleich2022}%
  \BibitemOpen
  \bibfield  {author} {\bibinfo {author} {\bibfnamefont {P.}~\bibnamefont {Schleich}}, \bibinfo {author} {\bibfnamefont {J.~S.}\ \bibnamefont {Kottmann}},\ and\ \bibinfo {author} {\bibfnamefont {A.}~\bibnamefont {Aspuru-Guzik}},\ }\bibfield  {title} {\bibinfo {title} {Improving the accuracy of the variational quantum eigensolver for molecular systems by the explicitly-correlated perturbative [2]r12-correction},\ }\href {https://doi.org/10.1039/D2CP00247G} {\bibfield  {journal} {\bibinfo  {journal} {Phys. Chem. Chem. Phys.}\ }\textbf {\bibinfo {volume} {24}},\ \bibinfo {pages} {13550} (\bibinfo {year} {2022})}\BibitemShut {NoStop}%
\bibitem [{\citenamefont {McArdle}\ and\ \citenamefont {Tew}(2020)}]{McArdle2020}%
  \BibitemOpen
  \bibfield  {author} {\bibinfo {author} {\bibfnamefont {S.}~\bibnamefont {McArdle}}\ and\ \bibinfo {author} {\bibfnamefont {D.~P.}\ \bibnamefont {Tew}},\ }\bibfield  {title} {\bibinfo {title} {Improving the accuracy of quantum computational chemistry using the transcorrelated method},\ }\href {https://arxiv.org/abs/2006.11181} {\bibfield  {journal} {\bibinfo  {journal} {arXiv preprint arXiv:2006.11181}\ } (\bibinfo {year} {2020})}\BibitemShut {NoStop}%
\bibitem [{\citenamefont {Kumar}\ \emph {et~al.}(2022)\citenamefont {Kumar}, \citenamefont {Asthana}, \citenamefont {Masteran}, \citenamefont {Valeev}, \citenamefont {Zhang}, \citenamefont {Cincio}, \citenamefont {Tretiak},\ and\ \citenamefont {Dub}}]{Kumar2022}%
  \BibitemOpen
  \bibfield  {author} {\bibinfo {author} {\bibfnamefont {A.}~\bibnamefont {Kumar}}, \bibinfo {author} {\bibfnamefont {A.}~\bibnamefont {Asthana}}, \bibinfo {author} {\bibfnamefont {C.}~\bibnamefont {Masteran}}, \bibinfo {author} {\bibfnamefont {E.~F.}\ \bibnamefont {Valeev}}, \bibinfo {author} {\bibfnamefont {Y.}~\bibnamefont {Zhang}}, \bibinfo {author} {\bibfnamefont {L.}~\bibnamefont {Cincio}}, \bibinfo {author} {\bibfnamefont {S.}~\bibnamefont {Tretiak}},\ and\ \bibinfo {author} {\bibfnamefont {P.~A.}\ \bibnamefont {Dub}},\ }\bibfield  {title} {\bibinfo {title} {Quantum simulation of molecular electronic states with a transcorrelated hamiltonian: Higher accuracy with fewer qubits},\ }\href {https://doi.org/10.1021/acs.jctc.2c00520} {\bibfield  {journal} {\bibinfo  {journal} {J. Chem. Theory Comput.}\ }\textbf {\bibinfo {volume} {18}},\ \bibinfo {pages} {5312} (\bibinfo {year} {2022})}\BibitemShut {NoStop}%
\bibitem [{\citenamefont {Sokolov}\ \emph {et~al.}(2023)\citenamefont {Sokolov}, \citenamefont {Dobrautz}, \citenamefont {Luo}, \citenamefont {Alavi},\ and\ \citenamefont {Tavernelli}}]{Sokolov2023}%
  \BibitemOpen
  \bibfield  {author} {\bibinfo {author} {\bibfnamefont {I.~O.}\ \bibnamefont {Sokolov}}, \bibinfo {author} {\bibfnamefont {W.}~\bibnamefont {Dobrautz}}, \bibinfo {author} {\bibfnamefont {H.}~\bibnamefont {Luo}}, \bibinfo {author} {\bibfnamefont {A.}~\bibnamefont {Alavi}},\ and\ \bibinfo {author} {\bibfnamefont {I.}~\bibnamefont {Tavernelli}},\ }\bibfield  {title} {\bibinfo {title} {Orders of magnitude increased accuracy for quantum many-body problems on quantum computers via an exact transcorrelated method},\ }\href {https://doi.org/10.1103/PhysRevResearch.5.023174} {\bibfield  {journal} {\bibinfo  {journal} {Phys. Rev. Res.}\ }\textbf {\bibinfo {volume} {5}},\ \bibinfo {pages} {023174} (\bibinfo {year} {2023})}\BibitemShut {NoStop}%
\bibitem [{\citenamefont {Huggins}\ \emph {et~al.}(2022)\citenamefont {Huggins}, \citenamefont {O’Gorman}, \citenamefont {Rubin}, \citenamefont {Reichman}, \citenamefont {Babbush},\ and\ \citenamefont {Lee}}]{Huggins2022}%
  \BibitemOpen
  \bibfield  {author} {\bibinfo {author} {\bibfnamefont {W.~J.}\ \bibnamefont {Huggins}}, \bibinfo {author} {\bibfnamefont {B.~A.}\ \bibnamefont {O’Gorman}}, \bibinfo {author} {\bibfnamefont {N.~C.}\ \bibnamefont {Rubin}}, \bibinfo {author} {\bibfnamefont {D.~R.}\ \bibnamefont {Reichman}}, \bibinfo {author} {\bibfnamefont {R.}~\bibnamefont {Babbush}},\ and\ \bibinfo {author} {\bibfnamefont {J.}~\bibnamefont {Lee}},\ }\bibfield  {title} {\bibinfo {title} {Unbiasing fermionic quantum monte carlo with a quantum computer},\ }\href {https://doi.org/10.1038/s41586-021-04351-z} {\bibfield  {journal} {\bibinfo  {journal} {Nature}\ }\textbf {\bibinfo {volume} {603}},\ \bibinfo {pages} {416} (\bibinfo {year} {2022})}\BibitemShut {NoStop}%
\bibitem [{\citenamefont {Zhang}\ and\ \citenamefont {Krakauer}(2003)}]{Zhang2003}%
  \BibitemOpen
  \bibfield  {author} {\bibinfo {author} {\bibfnamefont {S.}~\bibnamefont {Zhang}}\ and\ \bibinfo {author} {\bibfnamefont {H.}~\bibnamefont {Krakauer}},\ }\bibfield  {title} {\bibinfo {title} {Quantum monte carlo method using phase-free random walks with slater determinants},\ }\href {https://doi.org/10.1103/PhysRevLett.90.136401} {\bibfield  {journal} {\bibinfo  {journal} {Phys. Rev. Lett.}\ }\textbf {\bibinfo {volume} {90}},\ \bibinfo {pages} {136401} (\bibinfo {year} {2003})}\BibitemShut {NoStop}%
\bibitem [{\citenamefont {Zhao}\ \emph {et~al.}(2025)\citenamefont {Zhao}, \citenamefont {Goings}, \citenamefont {Aboumrad}, \citenamefont {Arrasmith}, \citenamefont {Calderin}, \citenamefont {Churchill}, \citenamefont {Gabay}, \citenamefont {Harvey-Brown}, \citenamefont {Hiles}, \citenamefont {Kaja} \emph {et~al.}}]{Zhao2025arXiv}%
  \BibitemOpen
  \bibfield  {author} {\bibinfo {author} {\bibfnamefont {L.}~\bibnamefont {Zhao}}, \bibinfo {author} {\bibfnamefont {J.~J.}\ \bibnamefont {Goings}}, \bibinfo {author} {\bibfnamefont {W.}~\bibnamefont {Aboumrad}}, \bibinfo {author} {\bibfnamefont {A.}~\bibnamefont {Arrasmith}}, \bibinfo {author} {\bibfnamefont {L.}~\bibnamefont {Calderin}}, \bibinfo {author} {\bibfnamefont {S.}~\bibnamefont {Churchill}}, \bibinfo {author} {\bibfnamefont {D.}~\bibnamefont {Gabay}}, \bibinfo {author} {\bibfnamefont {T.}~\bibnamefont {Harvey-Brown}}, \bibinfo {author} {\bibfnamefont {M.}~\bibnamefont {Hiles}}, \bibinfo {author} {\bibfnamefont {M.}~\bibnamefont {Kaja}}, \emph {et~al.},\ }\bibfield  {title} {\bibinfo {title} {Quantum-classical auxiliary field quantum monte carlo with matchgate shadows on trapped ion quantum computers},\ }\href {https://arxiv.org/abs/2506.22408} {\bibfield  {journal} {\bibinfo  {journal} {arXiv preprint arXiv:2506.22408}\ } (\bibinfo {year} {2025})}\BibitemShut {NoStop}%
\bibitem [{\citenamefont {Goings}\ \emph {et~al.}(2025)\citenamefont {Goings}, \citenamefont {Shin}, \citenamefont {Noh}, \citenamefont {Kyoung}, \citenamefont {Kim}, \citenamefont {Baek}, \citenamefont {Roetteler}, \citenamefont {Epifanovsky},\ and\ \citenamefont {Zhao}}]{Goings2025arXiv}%
  \BibitemOpen
  \bibfield  {author} {\bibinfo {author} {\bibfnamefont {J.~J.}\ \bibnamefont {Goings}}, \bibinfo {author} {\bibfnamefont {K.}~\bibnamefont {Shin}}, \bibinfo {author} {\bibfnamefont {S.}~\bibnamefont {Noh}}, \bibinfo {author} {\bibfnamefont {W.}~\bibnamefont {Kyoung}}, \bibinfo {author} {\bibfnamefont {D.}~\bibnamefont {Kim}}, \bibinfo {author} {\bibfnamefont {J.}~\bibnamefont {Baek}}, \bibinfo {author} {\bibfnamefont {M.}~\bibnamefont {Roetteler}}, \bibinfo {author} {\bibfnamefont {E.}~\bibnamefont {Epifanovsky}},\ and\ \bibinfo {author} {\bibfnamefont {L.}~\bibnamefont {Zhao}},\ }\bibfield  {title} {\bibinfo {title} {Molecular properties in quantum-classical auxiliary-field quantum monte carlo: Correlated sampling with application to accurate nuclear forces},\ }\href {https://arxiv.org/abs/2507.17992} {\bibfield  {journal} {\bibinfo  {journal} {arXiv preprint arXiv:2507.17992}\ } (\bibinfo {year} {2025})}\BibitemShut {NoStop}%
\bibitem [{\citenamefont {Motta}\ and\ \citenamefont {Zhang}(2018)}]{Motta2018}%
  \BibitemOpen
  \bibfield  {author} {\bibinfo {author} {\bibfnamefont {M.}~\bibnamefont {Motta}}\ and\ \bibinfo {author} {\bibfnamefont {S.}~\bibnamefont {Zhang}},\ }\bibfield  {title} {\bibinfo {title} {Ab initio computations of molecular systems by the auxiliary-field quantum monte carlo method},\ }\href {https://doi.org/https://doi.org/10.1002/wcms.1364} {\bibfield  {journal} {\bibinfo  {journal} {WIREs Comput. Mol. Sci.}\ }\textbf {\bibinfo {volume} {8}},\ \bibinfo {pages} {e1364} (\bibinfo {year} {2018})}\BibitemShut {NoStop}%
\bibitem [{\citenamefont {Lee}\ \emph {et~al.}(2022{\natexlab{a}})\citenamefont {Lee}, \citenamefont {Pham},\ and\ \citenamefont {Reichman}}]{Lee2022}%
  \BibitemOpen
  \bibfield  {author} {\bibinfo {author} {\bibfnamefont {J.}~\bibnamefont {Lee}}, \bibinfo {author} {\bibfnamefont {H.~Q.}\ \bibnamefont {Pham}},\ and\ \bibinfo {author} {\bibfnamefont {D.~R.}\ \bibnamefont {Reichman}},\ }\bibfield  {title} {\bibinfo {title} {Twenty years of auxiliary-field quantum monte carlo in quantum chemistry: An overview and assessment on main group chemistry and bond-breaking},\ }\href {https://doi.org/10.1021/acs.jctc.2c00802} {\bibfield  {journal} {\bibinfo  {journal} {J. Chem. Theory Comput.}\ }\textbf {\bibinfo {volume} {18}},\ \bibinfo {pages} {7024} (\bibinfo {year} {2022}{\natexlab{a}})}\BibitemShut {NoStop}%
\bibitem [{\citenamefont {Chen}\ \emph {et~al.}(2023)\citenamefont {Chen}, \citenamefont {Zhang}, \citenamefont {E},\ and\ \citenamefont {Car}}]{Chen2023}%
  \BibitemOpen
  \bibfield  {author} {\bibinfo {author} {\bibfnamefont {Y.}~\bibnamefont {Chen}}, \bibinfo {author} {\bibfnamefont {L.}~\bibnamefont {Zhang}}, \bibinfo {author} {\bibfnamefont {W.}~\bibnamefont {E}},\ and\ \bibinfo {author} {\bibfnamefont {R.}~\bibnamefont {Car}},\ }\bibfield  {title} {\bibinfo {title} {Hybrid auxiliary field quantum monte carlo for molecular systems},\ }\href {https://doi.org/10.1021/acs.jctc.3c00038} {\bibfield  {journal} {\bibinfo  {journal} {J. Chem. Theory Comput.}\ }\textbf {\bibinfo {volume} {19}},\ \bibinfo {pages} {4484} (\bibinfo {year} {2023})}\BibitemShut {NoStop}%
\bibitem [{\citenamefont {Yoshida}\ \emph {et~al.}(2025)\citenamefont {Yoshida}, \citenamefont {Erhart}, \citenamefont {Murokoshi}, \citenamefont {Nakagawa}, \citenamefont {Mori}, \citenamefont {Tagami},\ and\ \citenamefont {Mizukami}}]{yoshida2025auxiliaryfieldquantummontecarlo}%
  \BibitemOpen
  \bibfield  {author} {\bibinfo {author} {\bibfnamefont {Y.}~\bibnamefont {Yoshida}}, \bibinfo {author} {\bibfnamefont {L.}~\bibnamefont {Erhart}}, \bibinfo {author} {\bibfnamefont {T.}~\bibnamefont {Murokoshi}}, \bibinfo {author} {\bibfnamefont {R.}~\bibnamefont {Nakagawa}}, \bibinfo {author} {\bibfnamefont {C.}~\bibnamefont {Mori}}, \bibinfo {author} {\bibfnamefont {H.}~\bibnamefont {Tagami}},\ and\ \bibinfo {author} {\bibfnamefont {W.}~\bibnamefont {Mizukami}},\ }\bibfield  {title} {\bibinfo {title} {Auxiliary-field quantum monte carlo method with seniority-zero trial wave function},\ }\href {https://arxiv.org/abs/2501.18937} {\bibfield  {journal} {\bibinfo  {journal} {arXiv preprint arXiv:2501.18937}\ } (\bibinfo {year} {2025})}\BibitemShut {NoStop}%
\bibitem [{\citenamefont {Kiser}\ \emph {et~al.}(2024)\citenamefont {Kiser}, \citenamefont {Beuerle},\ and\ \citenamefont {Simkovic}}]{kiser2024contextualsubspaceauxiliaryfieldquantum}%
  \BibitemOpen
  \bibfield  {author} {\bibinfo {author} {\bibfnamefont {M.}~\bibnamefont {Kiser}}, \bibinfo {author} {\bibfnamefont {M.}~\bibnamefont {Beuerle}},\ and\ \bibinfo {author} {\bibfnamefont {F.}~\bibnamefont {Simkovic}},\ }\bibfield  {title} {\bibinfo {title} {Contextual subspace auxiliary-field quantum monte carlo: Improved bias with reduced quantum resources},\ }\href {https://arxiv.org/abs/2408.06160} {\bibfield  {journal} {\bibinfo  {journal} {arXiv preprint arXiv:2408.06160}\ } (\bibinfo {year} {2024})}\BibitemShut {NoStop}%
\bibitem [{\citenamefont {Huang}\ \emph {et~al.}(2020)\citenamefont {Huang}, \citenamefont {Kueng},\ and\ \citenamefont {Preskill}}]{Huang2020}%
  \BibitemOpen
  \bibfield  {author} {\bibinfo {author} {\bibfnamefont {H.-Y.}\ \bibnamefont {Huang}}, \bibinfo {author} {\bibfnamefont {R.}~\bibnamefont {Kueng}},\ and\ \bibinfo {author} {\bibfnamefont {J.}~\bibnamefont {Preskill}},\ }\bibfield  {title} {\bibinfo {title} {Predicting many properties of a quantum system from very few measurements},\ }\href {https://doi.org/10.1038/s41567-020-0932-7} {\bibfield  {journal} {\bibinfo  {journal} {Nat. Phys.}\ }\textbf {\bibinfo {volume} {16}},\ \bibinfo {pages} {1050} (\bibinfo {year} {2020})}\BibitemShut {NoStop}%
\bibitem [{\citenamefont {Wan}\ \emph {et~al.}(2023)\citenamefont {Wan}, \citenamefont {Huggins}, \citenamefont {Lee},\ and\ \citenamefont {Babbush}}]{Wan2023}%
  \BibitemOpen
  \bibfield  {author} {\bibinfo {author} {\bibfnamefont {K.}~\bibnamefont {Wan}}, \bibinfo {author} {\bibfnamefont {W.~J.}\ \bibnamefont {Huggins}}, \bibinfo {author} {\bibfnamefont {J.}~\bibnamefont {Lee}},\ and\ \bibinfo {author} {\bibfnamefont {R.}~\bibnamefont {Babbush}},\ }\bibfield  {title} {\bibinfo {title} {Matchgate shadows for fermionic quantum simulation},\ }\href {https://doi.org/10.1007/s00220-023-04844-0} {\bibfield  {journal} {\bibinfo  {journal} {Commun. Math. Phys.}\ }\textbf {\bibinfo {volume} {404}},\ \bibinfo {pages} {629} (\bibinfo {year} {2023})}\BibitemShut {NoStop}%
\bibitem [{\citenamefont {Huang}\ \emph {et~al.}(2024)\citenamefont {Huang}, \citenamefont {Chen}, \citenamefont {Gupt}, \citenamefont {Suchara}, \citenamefont {Tran}, \citenamefont {McArdle},\ and\ \citenamefont {Galli}}]{huang2024evaluatingquantumclassicalquantummonte}%
  \BibitemOpen
  \bibfield  {author} {\bibinfo {author} {\bibfnamefont {B.}~\bibnamefont {Huang}}, \bibinfo {author} {\bibfnamefont {Y.-T.}\ \bibnamefont {Chen}}, \bibinfo {author} {\bibfnamefont {B.}~\bibnamefont {Gupt}}, \bibinfo {author} {\bibfnamefont {M.}~\bibnamefont {Suchara}}, \bibinfo {author} {\bibfnamefont {A.}~\bibnamefont {Tran}}, \bibinfo {author} {\bibfnamefont {S.}~\bibnamefont {McArdle}},\ and\ \bibinfo {author} {\bibfnamefont {G.}~\bibnamefont {Galli}},\ }\bibfield  {title} {\bibinfo {title} {Evaluating a quantum-classical quantum monte carlo algorithm with matchgate shadows},\ }\href {https://doi.org/10.1103/PhysRevResearch.6.043063} {\bibfield  {journal} {\bibinfo  {journal} {Phys. Rev. Res.}\ }\textbf {\bibinfo {volume} {6}},\ \bibinfo {pages} {043063} (\bibinfo {year} {2024})}\BibitemShut {NoStop}%
\bibitem [{\citenamefont {Takemori}\ \emph {et~al.}(2023)\citenamefont {Takemori}, \citenamefont {Teranishi}, \citenamefont {Mizukami},\ and\ \citenamefont {Yoshioka}}]{takemori2023balancing}%
  \BibitemOpen
  \bibfield  {author} {\bibinfo {author} {\bibfnamefont {N.}~\bibnamefont {Takemori}}, \bibinfo {author} {\bibfnamefont {Y.}~\bibnamefont {Teranishi}}, \bibinfo {author} {\bibfnamefont {W.}~\bibnamefont {Mizukami}},\ and\ \bibinfo {author} {\bibfnamefont {N.}~\bibnamefont {Yoshioka}},\ }\bibfield  {title} {\bibinfo {title} {Balancing error budget for fermionic k-rdm estimation},\ }\href {https://arxiv.org/abs/2312.17452} {\bibfield  {journal} {\bibinfo  {journal} {arXiv preprint arXiv:2312.17452}\ } (\bibinfo {year} {2023})}\BibitemShut {NoStop}%
\bibitem [{\citenamefont {Mazzola}\ and\ \citenamefont {Carleo}(2022)}]{mazzola2022exponentialchallengesunbiasingquantum}%
  \BibitemOpen
  \bibfield  {author} {\bibinfo {author} {\bibfnamefont {G.}~\bibnamefont {Mazzola}}\ and\ \bibinfo {author} {\bibfnamefont {G.}~\bibnamefont {Carleo}},\ }\bibfield  {title} {\bibinfo {title} {Exponential challenges in unbiasing quantum monte carlo algorithms with quantum computers},\ }\href {https://arxiv.org/abs/2205.09203} {\bibfield  {journal} {\bibinfo  {journal} {arXiv preprint arXiv:2205.09203}\ } (\bibinfo {year} {2022})}\BibitemShut {NoStop}%
\bibitem [{\citenamefont {Lee}\ \emph {et~al.}(2022{\natexlab{b}})\citenamefont {Lee}, \citenamefont {Reichman}, \citenamefont {Babbush}, \citenamefont {Rubin}, \citenamefont {Malone}, \citenamefont {O'Gorman},\ and\ \citenamefont {Huggins}}]{lee2022responseexponentialchallengesunbiasing}%
  \BibitemOpen
  \bibfield  {author} {\bibinfo {author} {\bibfnamefont {J.}~\bibnamefont {Lee}}, \bibinfo {author} {\bibfnamefont {D.~R.}\ \bibnamefont {Reichman}}, \bibinfo {author} {\bibfnamefont {R.}~\bibnamefont {Babbush}}, \bibinfo {author} {\bibfnamefont {N.~C.}\ \bibnamefont {Rubin}}, \bibinfo {author} {\bibfnamefont {F.~D.}\ \bibnamefont {Malone}}, \bibinfo {author} {\bibfnamefont {B.}~\bibnamefont {O'Gorman}},\ and\ \bibinfo {author} {\bibfnamefont {W.~J.}\ \bibnamefont {Huggins}},\ }\bibfield  {title} {\bibinfo {title} {Response to "exponential challenges in unbiasing quantum monte carlo algorithms with quantum computers"},\ }\href {https://arxiv.org/abs/2207.13776} {\bibfield  {journal} {\bibinfo  {journal} {arXiv preprint arXiv:2207.13776}\ } (\bibinfo {year} {2022}{\natexlab{b}})}\BibitemShut {NoStop}%
\bibitem [{\citenamefont {Kohda}\ \emph {et~al.}(2022)\citenamefont {Kohda}, \citenamefont {Imai}, \citenamefont {Kanno}, \citenamefont {Mitarai}, \citenamefont {Mizukami},\ and\ \citenamefont {Nakagawa}}]{Kohda2022}%
  \BibitemOpen
  \bibfield  {author} {\bibinfo {author} {\bibfnamefont {M.}~\bibnamefont {Kohda}}, \bibinfo {author} {\bibfnamefont {R.}~\bibnamefont {Imai}}, \bibinfo {author} {\bibfnamefont {K.}~\bibnamefont {Kanno}}, \bibinfo {author} {\bibfnamefont {K.}~\bibnamefont {Mitarai}}, \bibinfo {author} {\bibfnamefont {W.}~\bibnamefont {Mizukami}},\ and\ \bibinfo {author} {\bibfnamefont {Y.~O.}\ \bibnamefont {Nakagawa}},\ }\bibfield  {title} {\bibinfo {title} {Quantum expectation-value estimation by computational basis sampling},\ }\href {https://doi.org/10.1103/PhysRevResearch.4.033173} {\bibfield  {journal} {\bibinfo  {journal} {Phys. Rev. Res.}\ }\textbf {\bibinfo {volume} {4}},\ \bibinfo {pages} {033173} (\bibinfo {year} {2022})}\BibitemShut {NoStop}%
\bibitem [{\citenamefont {Bäumer}\ and\ \citenamefont {Woerner}(2024)}]{baumer2024measurementbasedlongrangeentanglinggates}%
  \BibitemOpen
  \bibfield  {author} {\bibinfo {author} {\bibfnamefont {E.}~\bibnamefont {Bäumer}}\ and\ \bibinfo {author} {\bibfnamefont {S.}~\bibnamefont {Woerner}},\ }\bibfield  {title} {\bibinfo {title} {Measurement-based long-range entangling gates in constant depth},\ }\href {https://arxiv.org/abs/2408.03064} {\bibfield  {journal} {\bibinfo  {journal} {arXiv preprint arXiv:2408.03064}\ } (\bibinfo {year} {2024})}\BibitemShut {NoStop}%
\bibitem [{\citenamefont {Erhart}\ \emph {et~al.}(2024)\citenamefont {Erhart}, \citenamefont {Yoshida}, \citenamefont {Khinevich},\ and\ \citenamefont {Mizukami}}]{Erhart2024}%
  \BibitemOpen
  \bibfield  {author} {\bibinfo {author} {\bibfnamefont {L.}~\bibnamefont {Erhart}}, \bibinfo {author} {\bibfnamefont {Y.}~\bibnamefont {Yoshida}}, \bibinfo {author} {\bibfnamefont {V.}~\bibnamefont {Khinevich}},\ and\ \bibinfo {author} {\bibfnamefont {W.}~\bibnamefont {Mizukami}},\ }\bibfield  {title} {\bibinfo {title} {Coupled cluster method tailored with quantum computing},\ }\href {https://doi.org/10.1103/PhysRevResearch.6.023230} {\bibfield  {journal} {\bibinfo  {journal} {Phys. Rev. Res.}\ }\textbf {\bibinfo {volume} {6}},\ \bibinfo {pages} {023230} (\bibinfo {year} {2024})}\BibitemShut {NoStop}%
\bibitem [{\citenamefont {Sugiyama}\ and\ \citenamefont {Koonin}(1986)}]{Sugiyama1986}%
  \BibitemOpen
  \bibfield  {author} {\bibinfo {author} {\bibfnamefont {G.}~\bibnamefont {Sugiyama}}\ and\ \bibinfo {author} {\bibfnamefont {S.}~\bibnamefont {Koonin}},\ }\bibfield  {title} {\bibinfo {title} {Auxiliary field monte-carlo for quantum many-body ground states},\ }\href {https://doi.org/10.1016/0003-4916(86)90107-7} {\bibfield  {journal} {\bibinfo  {journal} {Annals of Physics}\ }\textbf {\bibinfo {volume} {168}},\ \bibinfo {pages} {1} (\bibinfo {year} {1986})}\BibitemShut {NoStop}%
\bibitem [{\citenamefont {Malone}\ \emph {et~al.}(2023)\citenamefont {Malone}, \citenamefont {Mahajan}, \citenamefont {Spencer},\ and\ \citenamefont {Lee}}]{Malone2022}%
  \BibitemOpen
  \bibfield  {author} {\bibinfo {author} {\bibfnamefont {F.~D.}\ \bibnamefont {Malone}}, \bibinfo {author} {\bibfnamefont {A.}~\bibnamefont {Mahajan}}, \bibinfo {author} {\bibfnamefont {J.~S.}\ \bibnamefont {Spencer}},\ and\ \bibinfo {author} {\bibfnamefont {J.}~\bibnamefont {Lee}},\ }\bibfield  {title} {\bibinfo {title} {ipie: A python-based auxiliary-field quantum monte carlo program with flexibility and efficiency on cpus and gpus},\ }\href {https://doi.org/10.1021/acs.jctc.2c00934} {\bibfield  {journal} {\bibinfo  {journal} {J. Chem. Theory Comput.}\ }\textbf {\bibinfo {volume} {19}},\ \bibinfo {pages} {109} (\bibinfo {year} {2023})}\BibitemShut {NoStop}%
\bibitem [{che(2023)}]{chemqulacs}%
  \BibitemOpen
  \href@noop {} {\bibinfo {title} {Chemqulacs}},\ \bibinfo {howpublished} {\url{https://wmizukami.github.io/chemqulacs/}} (\bibinfo {year} {2023})\BibitemShut {NoStop}%
\bibitem [{\citenamefont {Suzuki}\ \emph {et~al.}(2021)\citenamefont {Suzuki}, \citenamefont {Kawase}, \citenamefont {Masumura}, \citenamefont {Hiraga}, \citenamefont {Nakadai}, \citenamefont {Chen}, \citenamefont {Nakanishi}, \citenamefont {Mitarai}, \citenamefont {Imai}, \citenamefont {Tamiya}, \citenamefont {Yamamoto}, \citenamefont {Yan}, \citenamefont {Kawakubo}, \citenamefont {Nakagawa}, \citenamefont {Ibe}, \citenamefont {Zhang}, \citenamefont {Yamashita}, \citenamefont {Yoshimura}, \citenamefont {Hayashi},\ and\ \citenamefont {Fujii}}]{Suzuki2021}%
  \BibitemOpen
  \bibfield  {author} {\bibinfo {author} {\bibfnamefont {Y.}~\bibnamefont {Suzuki}}, \bibinfo {author} {\bibfnamefont {Y.}~\bibnamefont {Kawase}}, \bibinfo {author} {\bibfnamefont {Y.}~\bibnamefont {Masumura}}, \bibinfo {author} {\bibfnamefont {Y.}~\bibnamefont {Hiraga}}, \bibinfo {author} {\bibfnamefont {M.}~\bibnamefont {Nakadai}}, \bibinfo {author} {\bibfnamefont {J.}~\bibnamefont {Chen}}, \bibinfo {author} {\bibfnamefont {K.~M.}\ \bibnamefont {Nakanishi}}, \bibinfo {author} {\bibfnamefont {K.}~\bibnamefont {Mitarai}}, \bibinfo {author} {\bibfnamefont {R.}~\bibnamefont {Imai}}, \bibinfo {author} {\bibfnamefont {S.}~\bibnamefont {Tamiya}}, \bibinfo {author} {\bibfnamefont {T.}~\bibnamefont {Yamamoto}}, \bibinfo {author} {\bibfnamefont {T.}~\bibnamefont {Yan}}, \bibinfo {author} {\bibfnamefont {T.}~\bibnamefont {Kawakubo}}, \bibinfo {author} {\bibfnamefont {Y.~O.}\ \bibnamefont {Nakagawa}}, \bibinfo {author} {\bibfnamefont {Y.}~\bibnamefont {Ibe}}, \bibinfo {author} {\bibfnamefont {Y.}~\bibnamefont {Zhang}},
  \bibinfo {author} {\bibfnamefont {H.}~\bibnamefont {Yamashita}}, \bibinfo {author} {\bibfnamefont {H.}~\bibnamefont {Yoshimura}}, \bibinfo {author} {\bibfnamefont {A.}~\bibnamefont {Hayashi}},\ and\ \bibinfo {author} {\bibfnamefont {K.}~\bibnamefont {Fujii}},\ }\bibfield  {title} {\bibinfo {title} {Qulacs: a fast and versatile quantum circuit simulator for research purpose},\ }\href {https://doi.org/10.22331/q-2021-10-06-559} {\bibfield  {journal} {\bibinfo  {journal} {Quantum}\ }\textbf {\bibinfo {volume} {5}},\ \bibinfo {pages} {559} (\bibinfo {year} {2021})}\BibitemShut {NoStop}%
\bibitem [{\citenamefont {Sun}\ \emph {et~al.}(2018)\citenamefont {Sun}, \citenamefont {Berkelbach}, \citenamefont {Blunt}, \citenamefont {Booth}, \citenamefont {Guo}, \citenamefont {Li}, \citenamefont {Liu}, \citenamefont {McClain}, \citenamefont {Sayfutyarova}, \citenamefont {Sharma}, \citenamefont {Wouters},\ and\ \citenamefont {Chan}}]{Sun2018}%
  \BibitemOpen
  \bibfield  {author} {\bibinfo {author} {\bibfnamefont {Q.}~\bibnamefont {Sun}}, \bibinfo {author} {\bibfnamefont {T.~C.}\ \bibnamefont {Berkelbach}}, \bibinfo {author} {\bibfnamefont {N.~S.}\ \bibnamefont {Blunt}}, \bibinfo {author} {\bibfnamefont {G.~H.}\ \bibnamefont {Booth}}, \bibinfo {author} {\bibfnamefont {S.}~\bibnamefont {Guo}}, \bibinfo {author} {\bibfnamefont {Z.}~\bibnamefont {Li}}, \bibinfo {author} {\bibfnamefont {J.}~\bibnamefont {Liu}}, \bibinfo {author} {\bibfnamefont {J.~D.}\ \bibnamefont {McClain}}, \bibinfo {author} {\bibfnamefont {E.~R.}\ \bibnamefont {Sayfutyarova}}, \bibinfo {author} {\bibfnamefont {S.}~\bibnamefont {Sharma}}, \bibinfo {author} {\bibfnamefont {S.}~\bibnamefont {Wouters}},\ and\ \bibinfo {author} {\bibfnamefont {G.~K.-L.}\ \bibnamefont {Chan}},\ }\bibfield  {title} {\bibinfo {title} {Pyscf: the python-based simulations of chemistry framework},\ }\href {https://doi.org/https://doi.org/10.1002/wcms.1340} {\bibfield  {journal} {\bibinfo  {journal} {WIREs Comput. Mol.
  Sci.}\ }\textbf {\bibinfo {volume} {8}},\ \bibinfo {pages} {e1340} (\bibinfo {year} {2018})}\BibitemShut {NoStop}%
\bibitem [{\citenamefont {Langhoff}\ and\ \citenamefont {Davidson}(1974)}]{Langhoff1974}%
  \BibitemOpen
  \bibfield  {author} {\bibinfo {author} {\bibfnamefont {S.~R.}\ \bibnamefont {Langhoff}}\ and\ \bibinfo {author} {\bibfnamefont {E.~R.}\ \bibnamefont {Davidson}},\ }\bibfield  {title} {\bibinfo {title} {Configuration interaction calculations on the nitrogen molecule},\ }\href {https://doi.org/10.1002/qua.560080106} {\bibfield  {journal} {\bibinfo  {journal} {Int. J. Quantum Chem.}\ }\textbf {\bibinfo {volume} {8}},\ \bibinfo {pages} {61} (\bibinfo {year} {1974})}\BibitemShut {NoStop}%
\bibitem [{\citenamefont {Neese}(2012)}]{Neese2012}%
  \BibitemOpen
  \bibfield  {author} {\bibinfo {author} {\bibfnamefont {F.}~\bibnamefont {Neese}},\ }\bibfield  {title} {\bibinfo {title} {The orca program system},\ }\href {https://doi.org/10.1002/wcms.81} {\bibfield  {journal} {\bibinfo  {journal} {WIREs Comput. Mol. Sci.}\ }\textbf {\bibinfo {volume} {2}},\ \bibinfo {pages} {73} (\bibinfo {year} {2012})}\BibitemShut {NoStop}%
\bibitem [{\citenamefont {Jewett}\ and\ \citenamefont {Bertozzi}(2010)}]{Jewett2010}%
  \BibitemOpen
  \bibfield  {author} {\bibinfo {author} {\bibfnamefont {J.~C.}\ \bibnamefont {Jewett}}\ and\ \bibinfo {author} {\bibfnamefont {C.~R.}\ \bibnamefont {Bertozzi}},\ }\bibfield  {title} {\bibinfo {title} {Cu-free click cycloaddition reactions in chemical biology},\ }\href {https://doi.org/10.1039/B901970G} {\bibfield  {journal} {\bibinfo  {journal} {Chem. Soc. Rev.}\ }\textbf {\bibinfo {volume} {39}},\ \bibinfo {pages} {1272} (\bibinfo {year} {2010})}\BibitemShut {NoStop}%
\bibitem [{\citenamefont {Devaraj}(2018)}]{devaraj_future_2018}%
  \BibitemOpen
  \bibfield  {author} {\bibinfo {author} {\bibfnamefont {N.~K.}\ \bibnamefont {Devaraj}},\ }\bibfield  {title} {\bibinfo {title} {The {Future} of {Bioorthogonal} {Chemistry}},\ }\href {https://doi.org/10.1021/acscentsci.8b00251} {\bibfield  {journal} {\bibinfo  {journal} {ACS Cent. Sci.}\ }\textbf {\bibinfo {volume} {4}},\ \bibinfo {pages} {952} (\bibinfo {year} {2018})},\ \bibinfo {note} {publisher: American Chemical Society}\BibitemShut {NoStop}%
\bibitem [{\citenamefont {Ess}\ and\ \citenamefont {Houk}(2005)}]{Ess2005}%
  \BibitemOpen
  \bibfield  {author} {\bibinfo {author} {\bibfnamefont {D.~H.}\ \bibnamefont {Ess}}\ and\ \bibinfo {author} {\bibfnamefont {K.~N.}\ \bibnamefont {Houk}},\ }\bibfield  {title} {\bibinfo {title} {Activation energies of pericyclic reactions: Performance of dft, mp2, and cbs-qb3 methods for the prediction of activation barriers and reaction energetics of 1,3-dipolar cycloadditions, and revised activation enthalpies for a standard set of hydrocarbon pericyclic reactions},\ }\href {https://doi.org/10.1021/jp052504v} {\bibfield  {journal} {\bibinfo  {journal} {J. Phys. Chem. A}\ }\textbf {\bibinfo {volume} {109}},\ \bibinfo {pages} {9542} (\bibinfo {year} {2005})}\BibitemShut {NoStop}%
\bibitem [{\citenamefont {Karton}\ and\ \citenamefont {Goerigk}(2015)}]{Karton2015}%
  \BibitemOpen
  \bibfield  {author} {\bibinfo {author} {\bibfnamefont {A.}~\bibnamefont {Karton}}\ and\ \bibinfo {author} {\bibfnamefont {L.}~\bibnamefont {Goerigk}},\ }\bibfield  {title} {\bibinfo {title} {Accurate reaction barrier heights of pericyclic reactions: Surprisingly large deviations for the cbs-qb3 composite method and their consequences in dft benchmark studies},\ }\href {https://doi.org/10.1002/jcc.23837} {\bibfield  {journal} {\bibinfo  {journal} {J. Comput. Chem.}\ }\textbf {\bibinfo {volume} {36}},\ \bibinfo {pages} {622} (\bibinfo {year} {2015})}\BibitemShut {NoStop}%
\bibitem [{\citenamefont {Stuyver}\ \emph {et~al.}(2023)\citenamefont {Stuyver}, \citenamefont {Jorner},\ and\ \citenamefont {Coley}}]{Stuyver2023}%
  \BibitemOpen
  \bibfield  {author} {\bibinfo {author} {\bibfnamefont {T.}~\bibnamefont {Stuyver}}, \bibinfo {author} {\bibfnamefont {K.}~\bibnamefont {Jorner}},\ and\ \bibinfo {author} {\bibfnamefont {C.~W.}\ \bibnamefont {Coley}},\ }\bibfield  {title} {\bibinfo {title} {Reaction profiles for quantum chemistry-computed [3+2] cycloaddition reactions},\ }\href {https://doi.org/10.1038/s41597-023-01977-8} {\bibfield  {journal} {\bibinfo  {journal} {Sci. Data}\ }\textbf {\bibinfo {volume} {10}},\ \bibinfo {pages} {66} (\bibinfo {year} {2023})}\BibitemShut {NoStop}%
\bibitem [{\citenamefont {Vermeeren}\ \emph {et~al.}(2022)\citenamefont {Vermeeren}, \citenamefont {Dalla~Tiezza}, \citenamefont {Wolf}, \citenamefont {Lahm}, \citenamefont {Allen}, \citenamefont {Schaefer}, \citenamefont {Hamlin},\ and\ \citenamefont {Bickelhaupt}}]{Vermeeren2022}%
  \BibitemOpen
  \bibfield  {author} {\bibinfo {author} {\bibfnamefont {P.}~\bibnamefont {Vermeeren}}, \bibinfo {author} {\bibfnamefont {M.}~\bibnamefont {Dalla~Tiezza}}, \bibinfo {author} {\bibfnamefont {M.~E.}\ \bibnamefont {Wolf}}, \bibinfo {author} {\bibfnamefont {M.~E.}\ \bibnamefont {Lahm}}, \bibinfo {author} {\bibfnamefont {W.~D.}\ \bibnamefont {Allen}}, \bibinfo {author} {\bibfnamefont {H.~F.}\ \bibnamefont {Schaefer}}, \bibinfo {author} {\bibfnamefont {T.~A.}\ \bibnamefont {Hamlin}},\ and\ \bibinfo {author} {\bibfnamefont {F.~M.}\ \bibnamefont {Bickelhaupt}},\ }\bibfield  {title} {\bibinfo {title} {Pericyclic reaction benchmarks: hierarchical computations targeting ccsdt(q)/cbs and analysis of dft performance},\ }\href {https://doi.org/10.1039/D2CP02234F} {\bibfield  {journal} {\bibinfo  {journal} {Phys. Chem. Chem. Phys.}\ }\textbf {\bibinfo {volume} {24}},\ \bibinfo {pages} {18028} (\bibinfo {year} {2022})}\BibitemShut {NoStop}%
\bibitem [{\citenamefont {Kanno}\ \emph {et~al.}(2023)\citenamefont {Kanno}, \citenamefont {Kohda}, \citenamefont {Imai}, \citenamefont {Koh}, \citenamefont {Mitarai}, \citenamefont {Mizukami},\ and\ \citenamefont {Nakagawa}}]{Kanno2023QSCI}%
  \BibitemOpen
  \bibfield  {author} {\bibinfo {author} {\bibfnamefont {K.}~\bibnamefont {Kanno}}, \bibinfo {author} {\bibfnamefont {M.}~\bibnamefont {Kohda}}, \bibinfo {author} {\bibfnamefont {R.}~\bibnamefont {Imai}}, \bibinfo {author} {\bibfnamefont {S.}~\bibnamefont {Koh}}, \bibinfo {author} {\bibfnamefont {K.}~\bibnamefont {Mitarai}}, \bibinfo {author} {\bibfnamefont {W.}~\bibnamefont {Mizukami}},\ and\ \bibinfo {author} {\bibfnamefont {Y.~O.}\ \bibnamefont {Nakagawa}},\ }\bibfield  {title} {\bibinfo {title} {Quantum-selected configuration interaction: classical diagonalization of hamiltonians in subspaces selected by quantum computers},\ }\href {https://arxiv.org/abs/2302.11320} {\bibfield  {journal} {\bibinfo  {journal} {arXiv preprint arXiv:2302.11320}\ } (\bibinfo {year} {2023})}\BibitemShut {NoStop}%
\bibitem [{\citenamefont {Nakagawa}\ \emph {et~al.}(2024)\citenamefont {Nakagawa}, \citenamefont {Kamoshita}, \citenamefont {Mizukami}, \citenamefont {Sudo},\ and\ \citenamefont {Ohnishi}}]{nakagawa2024adapt}%
  \BibitemOpen
  \bibfield  {author} {\bibinfo {author} {\bibfnamefont {Y.~O.}\ \bibnamefont {Nakagawa}}, \bibinfo {author} {\bibfnamefont {M.}~\bibnamefont {Kamoshita}}, \bibinfo {author} {\bibfnamefont {W.}~\bibnamefont {Mizukami}}, \bibinfo {author} {\bibfnamefont {S.}~\bibnamefont {Sudo}},\ and\ \bibinfo {author} {\bibfnamefont {Y.-y.}\ \bibnamefont {Ohnishi}},\ }\bibfield  {title} {\bibinfo {title} {Adapt-qsci: Adaptive construction of an input state for quantum-selected configuration interaction},\ }\href {https://doi.org/10.1021/acs.jctc.4c00846} {\bibfield  {journal} {\bibinfo  {journal} {J. Chem. Theory Comput.}\ }\textbf {\bibinfo {volume} {20}},\ \bibinfo {pages} {10817} (\bibinfo {year} {2024})}\BibitemShut {NoStop}%
\bibitem [{\citenamefont {Robledo-Moreno}\ \emph {et~al.}(2024)\citenamefont {Robledo-Moreno}, \citenamefont {Motta}, \citenamefont {Haas}, \citenamefont {Javadi-Abhari}, \citenamefont {Jurcevic}, \citenamefont {Kirby}, \citenamefont {Martiel}, \citenamefont {Sharma}, \citenamefont {Sharma}, \citenamefont {Shirakawa} \emph {et~al.}}]{robledo2024qsci}%
  \BibitemOpen
  \bibfield  {author} {\bibinfo {author} {\bibfnamefont {J.}~\bibnamefont {Robledo-Moreno}}, \bibinfo {author} {\bibfnamefont {M.}~\bibnamefont {Motta}}, \bibinfo {author} {\bibfnamefont {H.}~\bibnamefont {Haas}}, \bibinfo {author} {\bibfnamefont {A.}~\bibnamefont {Javadi-Abhari}}, \bibinfo {author} {\bibfnamefont {P.}~\bibnamefont {Jurcevic}}, \bibinfo {author} {\bibfnamefont {W.}~\bibnamefont {Kirby}}, \bibinfo {author} {\bibfnamefont {S.}~\bibnamefont {Martiel}}, \bibinfo {author} {\bibfnamefont {K.}~\bibnamefont {Sharma}}, \bibinfo {author} {\bibfnamefont {S.}~\bibnamefont {Sharma}}, \bibinfo {author} {\bibfnamefont {T.}~\bibnamefont {Shirakawa}}, \emph {et~al.},\ }\bibfield  {title} {\bibinfo {title} {Chemistry beyond exact solutions on a quantum-centric supercomputer},\ }\href {https://arxiv.org/abs/2405.05068} {\bibfield  {journal} {\bibinfo  {journal} {arXiv preprint arXiv:2405.05068}\ } (\bibinfo {year} {2024})}\BibitemShut {NoStop}%
\bibitem [{\citenamefont {Barison}\ \emph {et~al.}(2025)\citenamefont {Barison}, \citenamefont {Robledo~Moreno},\ and\ \citenamefont {Motta}}]{barison2024qsci}%
  \BibitemOpen
  \bibfield  {author} {\bibinfo {author} {\bibfnamefont {S.}~\bibnamefont {Barison}}, \bibinfo {author} {\bibfnamefont {J.}~\bibnamefont {Robledo~Moreno}},\ and\ \bibinfo {author} {\bibfnamefont {M.}~\bibnamefont {Motta}},\ }\bibfield  {title} {\bibinfo {title} {Quantum-centric computation of molecular excited states with extended sample-based quantum diagonalization},\ }\href {https://doi.org/10.1088/2058-9565/adb781} {\bibfield  {journal} {\bibinfo  {journal} {Quantum Sci. Technol.}\ }\textbf {\bibinfo {volume} {10}},\ \bibinfo {pages} {025034} (\bibinfo {year} {2025})}\BibitemShut {NoStop}%
\bibitem [{\citenamefont {Kaliakin}\ \emph {et~al.}(2024)\citenamefont {Kaliakin}, \citenamefont {Shajan}, \citenamefont {Moreno}, \citenamefont {Li}, \citenamefont {Mitra}, \citenamefont {Motta}, \citenamefont {Johnson}, \citenamefont {Saki}, \citenamefont {Das}, \citenamefont {Sitdikov} \emph {et~al.}}]{kaliakin2024qsci}%
  \BibitemOpen
  \bibfield  {author} {\bibinfo {author} {\bibfnamefont {D.}~\bibnamefont {Kaliakin}}, \bibinfo {author} {\bibfnamefont {A.}~\bibnamefont {Shajan}}, \bibinfo {author} {\bibfnamefont {J.~R.}\ \bibnamefont {Moreno}}, \bibinfo {author} {\bibfnamefont {Z.}~\bibnamefont {Li}}, \bibinfo {author} {\bibfnamefont {A.}~\bibnamefont {Mitra}}, \bibinfo {author} {\bibfnamefont {M.}~\bibnamefont {Motta}}, \bibinfo {author} {\bibfnamefont {C.}~\bibnamefont {Johnson}}, \bibinfo {author} {\bibfnamefont {A.~A.}\ \bibnamefont {Saki}}, \bibinfo {author} {\bibfnamefont {S.}~\bibnamefont {Das}}, \bibinfo {author} {\bibfnamefont {I.}~\bibnamefont {Sitdikov}}, \emph {et~al.},\ }\bibfield  {title} {\bibinfo {title} {Accurate quantum-centric simulations of supramolecular interactions},\ }\href {https://arxiv.org/abs/2410.09209} {\bibfield  {journal} {\bibinfo  {journal} {arXiv preprint arXiv:2410.09209}\ } (\bibinfo {year} {2024})}\BibitemShut {NoStop}%
\bibitem [{\citenamefont {Sugisaki}\ \emph {et~al.}(2024)\citenamefont {Sugisaki}, \citenamefont {Kanno}, \citenamefont {Itoko}, \citenamefont {Sakuma},\ and\ \citenamefont {Yamamoto}}]{sugisaki2024qsci}%
  \BibitemOpen
  \bibfield  {author} {\bibinfo {author} {\bibfnamefont {K.}~\bibnamefont {Sugisaki}}, \bibinfo {author} {\bibfnamefont {S.}~\bibnamefont {Kanno}}, \bibinfo {author} {\bibfnamefont {T.}~\bibnamefont {Itoko}}, \bibinfo {author} {\bibfnamefont {R.}~\bibnamefont {Sakuma}},\ and\ \bibinfo {author} {\bibfnamefont {N.}~\bibnamefont {Yamamoto}},\ }\bibfield  {title} {\bibinfo {title} {Hamiltonian simulation-based quantum-selected configuration interaction for large-scale electronic structure calculations with a quantum computer},\ }\href {https://arxiv.org/abs/2412.07218} {\bibfield  {journal} {\bibinfo  {journal} {arXiv preprint arXiv:2412.07218}\ } (\bibinfo {year} {2024})}\BibitemShut {NoStop}%
\bibitem [{\citenamefont {Mikkelsen}\ and\ \citenamefont {Nakagawa}(2024)}]{mikkelsen2024qsci}%
  \BibitemOpen
  \bibfield  {author} {\bibinfo {author} {\bibfnamefont {M.}~\bibnamefont {Mikkelsen}}\ and\ \bibinfo {author} {\bibfnamefont {Y.~O.}\ \bibnamefont {Nakagawa}},\ }\bibfield  {title} {\bibinfo {title} {Quantum-selected configuration interaction with time-evolved state},\ }\href {https://arxiv.org/abs/2412.13839} {\bibfield  {journal} {\bibinfo  {journal} {arXiv preprint arXiv:2412.13839}\ } (\bibinfo {year} {2024})}\BibitemShut {NoStop}%
\bibitem [{\citenamefont {Yu}\ \emph {et~al.}(2025)\citenamefont {Yu}, \citenamefont {Moreno}, \citenamefont {Iosue}, \citenamefont {Bertels}, \citenamefont {Claudino}, \citenamefont {Fuller}, \citenamefont {Groszkowski}, \citenamefont {Humble}, \citenamefont {Jurcevic}, \citenamefont {Kirby} \emph {et~al.}}]{yu2025qsci}%
  \BibitemOpen
  \bibfield  {author} {\bibinfo {author} {\bibfnamefont {J.}~\bibnamefont {Yu}}, \bibinfo {author} {\bibfnamefont {J.~R.}\ \bibnamefont {Moreno}}, \bibinfo {author} {\bibfnamefont {J.~T.}\ \bibnamefont {Iosue}}, \bibinfo {author} {\bibfnamefont {L.}~\bibnamefont {Bertels}}, \bibinfo {author} {\bibfnamefont {D.}~\bibnamefont {Claudino}}, \bibinfo {author} {\bibfnamefont {B.}~\bibnamefont {Fuller}}, \bibinfo {author} {\bibfnamefont {P.}~\bibnamefont {Groszkowski}}, \bibinfo {author} {\bibfnamefont {T.~S.}\ \bibnamefont {Humble}}, \bibinfo {author} {\bibfnamefont {P.}~\bibnamefont {Jurcevic}}, \bibinfo {author} {\bibfnamefont {W.}~\bibnamefont {Kirby}}, \emph {et~al.},\ }\bibfield  {title} {\bibinfo {title} {Quantum-centric algorithm for sample-based krylov diagonalization},\ }\href {https://arxiv.org/abs/2501.09702} {\bibfield  {journal} {\bibinfo  {journal} {arXiv preprint arXiv:2501.09702}\ } (\bibinfo {year} {2025})}\BibitemShut {NoStop}%
\bibitem [{\citenamefont {Kanno}\ \emph {et~al.}(2024)\citenamefont {Kanno}, \citenamefont {Nakamura}, \citenamefont {Kobayashi}, \citenamefont {Gocho}, \citenamefont {Hatanaka}, \citenamefont {Yamamoto},\ and\ \citenamefont {Gao}}]{Kanno2023QMC}%
  \BibitemOpen
  \bibfield  {author} {\bibinfo {author} {\bibfnamefont {S.}~\bibnamefont {Kanno}}, \bibinfo {author} {\bibfnamefont {H.}~\bibnamefont {Nakamura}}, \bibinfo {author} {\bibfnamefont {T.}~\bibnamefont {Kobayashi}}, \bibinfo {author} {\bibfnamefont {S.}~\bibnamefont {Gocho}}, \bibinfo {author} {\bibfnamefont {M.}~\bibnamefont {Hatanaka}}, \bibinfo {author} {\bibfnamefont {N.}~\bibnamefont {Yamamoto}},\ and\ \bibinfo {author} {\bibfnamefont {Q.}~\bibnamefont {Gao}},\ }\bibfield  {title} {\bibinfo {title} {Quantum computing quantum monte carlo with hybrid tensor network for electronic structure calculations},\ }\href {https://doi.org/10.1038/s41534-024-00851-8} {\bibfield  {journal} {\bibinfo  {journal} {npj Quantum Inf.}\ }\textbf {\bibinfo {volume} {10}},\ \bibinfo {pages} {56} (\bibinfo {year} {2024})}\BibitemShut {NoStop}%
\bibitem [{\citenamefont {Blunt}\ \emph {et~al.}(2024)\citenamefont {Blunt}, \citenamefont {Caune},\ and\ \citenamefont {Quiroz-Fernandez}}]{blunt2024quantumcomputingapproachfixednode}%
  \BibitemOpen
  \bibfield  {author} {\bibinfo {author} {\bibfnamefont {N.~S.}\ \bibnamefont {Blunt}}, \bibinfo {author} {\bibfnamefont {L.}~\bibnamefont {Caune}},\ and\ \bibinfo {author} {\bibfnamefont {J.}~\bibnamefont {Quiroz-Fernandez}},\ }\bibfield  {title} {\bibinfo {title} {A quantum computing approach to fixed-node monte carlo using classical shadows},\ }\href {https://arxiv.org/abs/2410.18901} {\bibfield  {journal} {\bibinfo  {journal} {arXiv preprint arXiv:2410.18901}\ } (\bibinfo {year} {2024})}\BibitemShut {NoStop}%
\bibitem [{\citenamefont {Halkier}\ \emph {et~al.}(1999)\citenamefont {Halkier}, \citenamefont {Helgaker}, \citenamefont {Jørgensen}, \citenamefont {Klopper},\ and\ \citenamefont {Olsen}}]{halkier_basis-set_1999}%
  \BibitemOpen
  \bibfield  {author} {\bibinfo {author} {\bibfnamefont {A.}~\bibnamefont {Halkier}}, \bibinfo {author} {\bibfnamefont {T.}~\bibnamefont {Helgaker}}, \bibinfo {author} {\bibfnamefont {P.}~\bibnamefont {Jørgensen}}, \bibinfo {author} {\bibfnamefont {W.}~\bibnamefont {Klopper}},\ and\ \bibinfo {author} {\bibfnamefont {J.}~\bibnamefont {Olsen}},\ }\bibfield  {title} {\bibinfo {title} {Basis-set convergence of the energy in molecular {Hartree}–{Fock} calculations},\ }\href {https://doi.org/https://doi.org/10.1016/S0009-2614(99)00179-7} {\bibfield  {journal} {\bibinfo  {journal} {Chem. Phys. Lett.}\ }\textbf {\bibinfo {volume} {302}},\ \bibinfo {pages} {437} (\bibinfo {year} {1999})}\BibitemShut {NoStop}%
\bibitem [{\citenamefont {Pansini}\ \emph {et~al.}(2016)\citenamefont {Pansini}, \citenamefont {Neto},\ and\ \citenamefont {Varandas}}]{Pansini2016}%
  \BibitemOpen
  \bibfield  {author} {\bibinfo {author} {\bibfnamefont {F.~N.~N.}\ \bibnamefont {Pansini}}, \bibinfo {author} {\bibfnamefont {A.~C.}\ \bibnamefont {Neto}},\ and\ \bibinfo {author} {\bibfnamefont {A.~J.~C.}\ \bibnamefont {Varandas}},\ }\bibfield  {title} {\bibinfo {title} {Extrapolation of hartree–fock and multiconfiguration self-consistent-field energies to the complete basis set limit},\ }\href {https://doi.org/10.1007/s00214-016-2016-4} {\bibfield  {journal} {\bibinfo  {journal} {Theor. Chem. Acc.}\ }\textbf {\bibinfo {volume} {135}},\ \bibinfo {pages} {261} (\bibinfo {year} {2016})}\BibitemShut {NoStop}%
\bibitem [{\citenamefont {Lesiuk}\ and\ \citenamefont {Jeziorski}(2019)}]{Lesiuk2019}%
  \BibitemOpen
  \bibfield  {author} {\bibinfo {author} {\bibfnamefont {M.}~\bibnamefont {Lesiuk}}\ and\ \bibinfo {author} {\bibfnamefont {B.}~\bibnamefont {Jeziorski}},\ }\bibfield  {title} {\bibinfo {title} {Complete basis set extrapolation of electronic correlation energies using the riemann zeta function},\ }\href {https://doi.org/10.1021/acs.jctc.9b00705} {\bibfield  {journal} {\bibinfo  {journal} {J. Chem. Theory Comput.}\ }\textbf {\bibinfo {volume} {15}},\ \bibinfo {pages} {5398} (\bibinfo {year} {2019})}\BibitemShut {NoStop}%
\bibitem [{\citenamefont {Helgaker}\ \emph {et~al.}(1997)\citenamefont {Helgaker}, \citenamefont {Klopper}, \citenamefont {Koch},\ and\ \citenamefont {Noga}}]{Helgaker1997}%
  \BibitemOpen
  \bibfield  {author} {\bibinfo {author} {\bibfnamefont {T.}~\bibnamefont {Helgaker}}, \bibinfo {author} {\bibfnamefont {W.}~\bibnamefont {Klopper}}, \bibinfo {author} {\bibfnamefont {H.}~\bibnamefont {Koch}},\ and\ \bibinfo {author} {\bibfnamefont {J.}~\bibnamefont {Noga}},\ }\bibfield  {title} {\bibinfo {title} {{Basis-set convergence of correlated calculations on water}},\ }\href {https://doi.org/10.1063/1.473863} {\bibfield  {journal} {\bibinfo  {journal} {J. Chem. Phys.}\ }\textbf {\bibinfo {volume} {106}},\ \bibinfo {pages} {9639} (\bibinfo {year} {1997})}\BibitemShut {NoStop}%
\bibitem [{\citenamefont {Amsler}\ \emph {et~al.}(2023)\citenamefont {Amsler}, \citenamefont {Deglmann}, \citenamefont {Degroote}, \citenamefont {Kaicher}, \citenamefont {Kiser}, \citenamefont {Kühn}, \citenamefont {Kumar}, \citenamefont {Maier}, \citenamefont {Samsonidze}, \citenamefont {Schroeder}, \citenamefont {Streif}, \citenamefont {Vodola}, \citenamefont {Wever},\ and\ \citenamefont {Group}}]{Amsler2023}%
  \BibitemOpen
  \bibfield  {author} {\bibinfo {author} {\bibfnamefont {M.}~\bibnamefont {Amsler}}, \bibinfo {author} {\bibfnamefont {P.}~\bibnamefont {Deglmann}}, \bibinfo {author} {\bibfnamefont {M.}~\bibnamefont {Degroote}}, \bibinfo {author} {\bibfnamefont {M.~P.}\ \bibnamefont {Kaicher}}, \bibinfo {author} {\bibfnamefont {M.}~\bibnamefont {Kiser}}, \bibinfo {author} {\bibfnamefont {M.}~\bibnamefont {Kühn}}, \bibinfo {author} {\bibfnamefont {C.}~\bibnamefont {Kumar}}, \bibinfo {author} {\bibfnamefont {A.}~\bibnamefont {Maier}}, \bibinfo {author} {\bibfnamefont {G.}~\bibnamefont {Samsonidze}}, \bibinfo {author} {\bibfnamefont {A.}~\bibnamefont {Schroeder}}, \bibinfo {author} {\bibfnamefont {M.}~\bibnamefont {Streif}}, \bibinfo {author} {\bibfnamefont {D.}~\bibnamefont {Vodola}}, \bibinfo {author} {\bibfnamefont {C.}~\bibnamefont {Wever}},\ and\ \bibinfo {author} {\bibfnamefont {Q.~M. S.~W.}\ \bibnamefont {Group}},\ }\bibfield  {title} {\bibinfo {title} {Classical and quantum trial wave functions in auxiliary-field
  quantum monte carlo applied to oxygen allotropes and a cubr2 model system},\ }\href {https://doi.org/10.1063/5.0146934} {\bibfield  {journal} {\bibinfo  {journal} {J. Chem. Phys.}\ }\textbf {\bibinfo {volume} {159}},\ \bibinfo {pages} {044119} (\bibinfo {year} {2023})}\BibitemShut {NoStop}%
\end{thebibliography}%

\end{document}